\documentclass[11pt]{article}
\usepackage{latexsym}
\usepackage{graphics,graphicx}
\usepackage{amsmath,amsfonts,amssymb}
\usepackage{epsfig}
\usepackage{lscape}
\usepackage{hyperref}
\def\NO{\nonumber}

\newcommand{\be}{\begin{equation}}
\newcommand{\ee}{\end{equation}}

\def\bea{\begin{eqnarray}}
\def\eea{\end{eqnarray}}

\def\beqx{\begin{displaymath}}
\def\eeqx{\end{displaymath}}

\newcommand{\bmat}{\left(\begin{array}}
\newcommand{\emat}{\end{array}\right)}

\def\a{\alpha}

\def\d{\delta}

\def\f{\phi}
\def\g{\gamma}
\def\h{\eta}

\def\k{\kappa}
\def\l{\lambda}
\def\m{\mu}
\def\n{\nu}

    \def\om{\omega}
\def\p{\pi}

    \def\th{\theta}
\def\r{\rho}
\def\s{\sigma}
\def\t{\tau}
\def\x{\xi}

\def\D{\Delta}

\def\G{\Gamma}

\def\L{\Lambda}

    \def\Om{\Omega}
\def\P{\Pi}

\def\S{\Sigma}

\def\ve{\varepsilon}

\def\vf{\varphi}

\def\ca{{\cal A}}

\def\cc{{\cal C}}
\def\cd{{\cal D}}

\def\ch{{\cal H}}
\def\ci{{\cal I}}

\def\cl{{\cal L}}
\def\cm{{\cal M}}
\def\cn{{\cal N}}
\def\co{{\cal O}}
\def\cp{{\cal P}}

\def\cs{{\cal S}}
\def\ct{{\cal T}}

\def\bo{{\raise-.3ex\hbox{\large$\Box$}}}               
\def\pa{\partial}                                       
\def\face{{\raise.2ex\hbox{$\displaystyle \bigodot$}\mskip-2.2mu \llap {$\ddot
        \smile$}}}                                   
\def\>{\rangle}                                      
\def\<{\langle}                                      

\newcommand{\sub}[1]{\phantom{}_{(#1)}\phantom{}}    
\def\leftrightarrowfill{$\mathsurround=0pt \mathord\leftarrow \mkern-6mu
        \cleaders\hbox{$\mkern-2mu \mathord- \mkern-2mu$}\hfill
        \mkern-6mu \mathord\rightarrow$}        
\def\dvec#1{\vbox{\ialign{##\crcr
        \leftrightarrowfill\crcr\noalign{\kern-1pt\nointerlineskip}
        $\hfil\displaystyle{#1}\hfil$\crcr}}}           

\def\-{\hphantom{-}}

\title{
\begin{minipage}{6.5in}
\flushright{\small CERN-PH-TH/2010-154}
\end{minipage}\\
\vskip 1.0in
Holographic renormalization as a canonical transformation
}

\author{Ioannis Papadimitriou\\ \\
{\it Department of Physics, CERN -- Theory Division,}\\ \\ {\it CH--1211 Geneva 23, Switzerland}\\ \\
e-mail: {\tt Ioannis.Papadimitriou@cern.ch}}

\date{}

\begin{document}
\renewcommand{\theequation}{\arabic{section}.\arabic{subsection}.\arabic{equation}}
\maketitle

\begin{abstract}

The gauge/string dualities have drawn attention to a class of variational problems on a boundary at infinity, which
are not well defined unless a certain boundary term is added to the classical action. In the context of supergravity
in asymptotically AdS spaces these problems are systematically addressed by the method of holographic renormalization.
We argue that this class of a priori ill defined variational problems extends far beyond the realm of holographic dualities.
As we show, exactly the same issues arise  in
gravity in non asymptotically AdS spaces, in point particles with certain unbounded from below potentials, and even
fundamental strings in flat or AdS backgrounds. We show that the variational problem in all such cases can be made
well defined by the following procedure, which is intrinsic to the system in question and does not rely on the
existence of a holographically dual theory: (i) The first step is the construction of the space of the most general
asymptotic solutions of the classical equations of motion that inherits a well defined symplectic form from that on
phase space. The requirement of a well defined symplectic form is essential and often leads to a necessary
repackaging of the degrees of freedom. (ii) Once the space of asymptotic solutions has been constructed in
terms of the correct degrees of freedom, then there exists a boundary term that is obtained as a certain
solution of the Hamilton-Jacobi equation which simultaneously makes the variational problem well defined and
preserves the symplectic form. This procedure is identical to holographic renormalization in the case of asymptotically
AdS gravity, but it is applicable to any Hamiltonian system.

\end{abstract}

\newpage

\tableofcontents
\addtocontents{toc}{\protect\setcounter{tocdepth}{3}}

\section{Introduction}

Holographic dualities have been at the center of some of the most fascinating developments in
theoretical physics in recent years, and holographic renormalization
\cite{Henningson:1998gx, Balasubramanian:1999re, Emparan:1999pm, Kraus:1999di,
de_Boer:1999xf,de_Haro:2000xn,Bianchi:2001kw,Martelli:2002sp,Papadimitriou:2004ap} was developed as an integral
part of the dictionary relating observables on the two sides of such dualities. This historical
perspective is probably behind the widespread belief that holographic renormalization is a
procedure that is applicable exclusively in the context of holographic dualities. In this article
we aim to demonstrate that this is not quite true.

We will see that the issues which holographic renormalization is designed to address arise in a wide class of Hamiltonian systems, with or without gravity, which do not necessarily admit an obvious holographically dual description. The ubiquitous
common feature of these systems is that the variational problem with time-independent initial/final conditions at past/future infinity along some suitable Hamiltonian `time' $t$ -- not necessarily the physical time -- is a priori ill defined. In order to make the variational problem well defined one needs to consider variations that asymptotically approach generic solutions of the equations of motion. The space of such asymptotic solutions must necessarily inherit a well defined symplectic form from that
on phase space. Note that this is a very non trivial requirement. It means that the asymptotic solutions must be sufficiently
general and often it forces us to reparameterize the asymptotic solutions in terms of different degrees of freedom.
Such a reparameterization of the degrees of freedom is generically mandatory when the asymptotic form of the solutions
depends on the Laplacian in the transverse directions. As we will see below, this happens in the case of a massless
scalar field in flat Euclidean space. In order to construct a space of asymptotic solutions that carries a well defined symplectic form one must decompose the scalar in terms of an infinite set of symmetric traceless tensors. Another example is Type IIB
supergravity on asymptotically $AdS_5\times S^5$ backgrounds. Again the asymptotic form of the solutions depends on the
Laplacian on the $S^5$ and so the construction of the space of asymptotic solutions with a well defined symplectic form
makes it mandatory to decompose the ten dimensional fields in $S^5$ harmonics. The Kaluza-Klein reduction, therefore, is
a consequence of the construction of a well defined space of asymptotic solutions. In particular, the boundary terms
that make the variational problem well defined {\em must} be formulated in terms of the KK fields and not the ten
dimensional fields.

Once the space of asymptotic solutions has been constructed, the variational problem in phase space can be mapped to one in the space of asymptotic solutions. This is achieved by adding a suitable boundary term on the initial/final surface, which is systematically derived as a certain solution of the Hamilton-Jacobi equation. It is crucial, however, that the Hamilton-Jacobi equation  -- and hence the boundary term -- be formulated in terms of the degrees of freedom that parameterize the space of asymptotic solutions carrying a well defined symplectic form. This ensures that the resulting boundary term not only makes
the variational problem well defined, but also preserves the symplectic form. As we will show, the addition of such a boundary
term amounts to a canonical transformation that asymptotically, i.e. as $t\to\infty$ or $t\to-\infty$, diagonalizes the symplectic map, $\varphi^*_t$, between the cotangent bundle of phase space and the cotangent bundle of the space of asymptotic solutions.
The diagonalization of this map is precisely what allows us to map the variational problem in phase space to one on
the space of asymptotic solutions.

It should be emphasized that making the classical variational problem well defined,
automatically makes the full quantum path integral well defined, which follows from the fact
that we are integrating over paths that asymptotically approach classical solutions and, hence, the WKB approximation is
asymptotically exact. In many respects, the above construction can be understood as computing the ground state
wavefunctional of the system. In gravity this amounts to computing the Hartle-Hawking wavefunction of the ground state
\cite{Hartle:1983ai}. A similar situation arises in the computation of the wavefunction corresponding to the identity operator
in Liouville theory when evaluating the path integral over the disc with no insertions \cite{Seiberg:1990eb}. In
Liouville theory this wavefunction is non normalizable and so it does not define a state, but even in CFTs where
this path integral does define a state in the Hilbert space, one needs to add a suitable boundary term in order
to define the path integral on a disc of infinite radius. We will see this explicitly in the last section, where we
work out this construction for the free boson CFT.

This procedure of making the variational problem well defined, which by a slight abuse of terminology we will
continue to call `holographic renormalization', is applicable in principle to any Hamiltonian system in non compact spaces.
The prescription given above is extremely general. Its successful implementation, however, is contingent upon the
construction of the space of asymptotic solutions with a non degenerate symplectic form, which is more difficult
in some systems than in others. Besides the systems we discuss below, we expect this construction to be applicable,
for example, to asymptotically flat gravity, higher dimensional supergravity in various backgrounds, D-branes with
non trivial embeddings, as well as strings in general backgrounds. Some of these systems will be discussed elsewhere
\cite{Papadimitriou-AF, Papadimitriou-HQCD}.

As we already mentioned, this procedure of making the variational problem well defined is intrinsic to the system whose
variational problem we are considering. In particular, it makes no reference to a holographically dual theory -- it
does not require the existence of a holographic duality, nor does it imply the existence of such a duality. However,
we claim that when such a holographic duality does exist, then making the variational problem of the `bulk' theory
well defined automatically leads to the correct observables of the holographically dual theory. This can be seen,
for example, from applying our procedure to Type IIB supergravity on asymptotically $AdS_5\times S^5$ backgrounds
\cite{Skenderis:2006uy}.
As we discussed above, constructing the space of asymptotic solutions with a well defined symplectic form forces
us to decompose the ten dimensional fields in terms of $S^5$ harmonics, leading to the correct spectrum of (BPS) operators
of $\cn=4$ super Yang-Mills. Moreover, the canonical transformation that is induced by the boundary term involving
the five dimensional fields leads to the AdS/CFT dictionary, where the fields induced on the boundary by the five
dimensional fields are identified with the sources of the dual operators, while their renormalized one point functions
are identified with the canonically transformed conjugate momenta.

The rest of the paper is organized as follows. In section \ref{toy} we argue that many of the essential features
of the general class of problems we want to address arise even in the simplistic context of a one dimensional point
particle in a certain class of potentials that are unbounded from below, yet allow the Hamiltonian to be self-adjoint,
such that there is still a well defined unitary time evolution operator. We use this toy example to demonstrate the connection between the variational problem at infinite future and canonical transformations, and to explicitly demonstrate the effect
of holographic renormalization on the quantum path integral. In section \ref{examples} we then discuss a number
of physically more interesting examples and make the connection between the boundary term required to make the
variational problem well defined and the canonical transformation that asymptotically diagonalizes the symplectic
map $\varphi_t^*$. The examples we consider are a scalar field in a fixed AdS or flat Euclidean background,
asymptotically AdS gravity, and finally strings in flat and anti de Sitter spaces.

\section{Holographic renormalization of point particles}
\label{toy}
\setcounter{equation}{0}

Consider a point particle in one space dimension described by the classical action
\be\label{particle-action}
S=\int_0^t dt' L=\int_0^t dt'\left(\frac12 \dot{q}^2-V(q)\right).
\ee
The Hamiltonian
\be\label{particle-Hamiltonian}
H=\frac12 p^2+V(q),
\ee
where $p=\dot{q}$, is conserved and hence the time it takes for a particle of energy $E$ to travel from $q_0$ at time $t_0$ to $q$
at time $t$ is given by
\be\label{time-integral}
t-t_0=\int_{q_0}^q \frac{dq'}{\sqrt{2(E-V(q'))}}.
\ee
We will be interested in potentials $V(q)$ for which the integral on the RHS diverges as $q\to\infty$, so that
the particle reaches infinity at infinite time. Quantum mechanically, this condition translates to the statement that the Hamiltonian does not require a self-adjoint extension \cite{Reed:1975uy, Carreau:1990is}. Importantly, in
this case, although the energy eigenstates are not $L^2$-normalizable, they are still  $\d$-function normalizable.\footnote{This
holds irrespectively of whether the potential goes to $-\infty$, a constant, or vanishes as $q\to\infty$, provided
the RHS of (\ref{time-integral}) diverges. For the case $V(q)\to-\infty$ specifically, which we will focus on below, the spectrum of the Hamiltonian is continuous and energy eigenfunctions, $\psi_E(q)$, satisfy
\be
\int^\infty dq \psi^*_E(q)\psi_{E'}(q)
\sim \int^\infty \frac{dq}{\sqrt{-2V(q)}}\exp\left(-i(E-E')\int^q \frac{dq'}{\sqrt{-2V(q')}}\right)\propto \d(E-E').
\ee}

\subsection{The variational problem}
\setcounter{equation}{0}

Let us now examine a generic variation of the action (\ref{particle-action}). We have,
\bea
\d S&=&-\int_0^t dt'\left(\ddot{q}+V'(q)\right)\d q+L\d t+p\d q\NO\\
&=&-\int_0^t dt'\left(\ddot{q}+V'(q)\right)\d q+(p\dot{q}-H)\d t+p \d q,
\eea
where we do not consider boundary terms at $t=0$ as they are irrelevant for
our present discussion.  We focus instead on the
boundary conditions at $t$. The usual variational problem is defined by keeping $t=t_0$ fixed and
$\d q=0$ at $t=t_0$. These boundary conditions though are not well defined when we want to consider the variational problem
for $t\in [0,\infty)$. Clearly, if we send $t\to\infty$ we cannot require that $\d t=0$. It follows that in order to
be able to impose the time independent Dirichlet boundary condition $\d q=0$ at $t=\infty$ we must add a boundary term,
$S_b(q(t))$, to the action. The variation of the total action will then be
\bea
\d (S+S_b)&=&-\int_0^t dt'\left(\ddot{q}+V'(q)\right)\d q+(p\dot{q}-H)\d t+p \d q +S'_b(q)(\d q+\dot{q}\d t)\NO\\
&=& -\int_0^t dt'\left(\ddot{q}+V'(q)\right)\d q+\left((p+S'_b(q))\dot{q}-H\right)\d t+(p+S'_b(q)) \d q.
\eea
The condition that determines $S_b(q)$ is that the coefficient of $\d t$ should vanish as $t\to\infty$. Namely,
\be\label{particle-renormalizability-condition}
(p+S'_b(q))\dot{q}-H=L+\dot{S}_b(q)=\frac{d}{dt}\left(S+S_b\right)\xrightarrow{t\to\infty} 0.
\ee
We therefore arrive at a general condition for the variational problem to be well defined at $t=\infty$:
The action (\ref{particle-action}) admits a time independent Dirichlet boundary condition at $t=\infty$ if and only if
 \begin{enumerate}
  \item the potential is such that generic solutions of the equations of motion satisfy $q(t)\xrightarrow{t\to\infty} \infty$,

  \item there exists an energy independent function $S_b(q)$ such that (\ref{particle-renormalizability-condition}) holds
  for any solution with $q(t)\xrightarrow{t\to\infty} \infty$.
 \end{enumerate}
We demand that $S_b(q)$ does not explicitly depend on the energy $E$ since the same $S_b(q)$ must ensure that
(\ref{particle-renormalizability-condition}) holds for all solutions satisfying $q(t)\xrightarrow{t\to\infty} \infty$. Moreover,
note that if there exists a boundary term $S_b$ that ensures that $|S+S_b|<\infty$ as $t\to\infty$, then condition
(\ref{particle-renormalizability-condition}) is automatically satisfied. This result is a direct demonstration of the connection between the finiteness of the on-shell action, which was the historical motivation for holographic renormalization, and the requirement that the variational problem at infinity be well defined, which was first observed in \cite{Papadimitriou:2005ii}
in the context of gravity in asymptotically locally AdS spaces. However, condition (\ref{particle-renormalizability-condition})
for the variational problem to be well defined is slightly weaker than the stronger condition that $S+S_b$ admit a finite limit
at infinity.\footnote{We are grateful to the referee for correctly pointing out to us that these two conditions are not
exactly equivalent.}

As we now show, the point particle example is simple enough to allow us to classify all actions (\ref{particle-action})
admitting time independent Dirichlet boundary conditions at $t=\infty$ and to determine explicitly the corresponding boundary term $S_b(q)$. In fact, as it turns out, all potentials $V(q)$ admit time independent Dirichlet boundary conditions at infinity provided they satisfy $V(q)\to -\infty$ as $q\to\infty$, but also $\int^qdq'1/\sqrt{-V(q')}\to \infty$ as $q\to\infty$. From energy conservation follows that
\be\label{particle-momentum}
\dot{q}=\pm\sqrt{2(E-V(q))}.
\ee
We want to consider generic solutions for which $q(t)\xrightarrow{t\to\infty} \infty$ and so we pick the plus sign for $\dot{q}$.
If $|V(q)|<\infty$ as $q\to\infty$, then there is no energy independent function $S_b(q)$ satisfying the condition
(\ref{particle-renormalizability-condition}) and so we consider only potentials such that $V(q)\to -\infty$ as $q\to\infty$.
Hence, for large $q$
\be
\dot{q}=\sqrt{-2V(q)}\left(1-\frac{E}{2V(q)}+\cdots\right).
\ee
Inserting this into (\ref{particle-renormalizability-condition}) we obtain
\be
\left(\sqrt{-2V(q)}+S_b'(q)\right)\left(\frac{E}{\sqrt{-2V(q)}}+\sqrt{-2V(q)}\right)\xrightarrow{q\to\infty} 0.
\ee
Therefore, an energy independent function $S_b(q)$ satisfying (\ref{particle-renormalizability-condition})
always exists for such potentials and it is given by
\be\label{particle-boundary-term}
S_b(q)=-\int^qdq'\sqrt{-2V(q')}.
\ee

\subsection{Canonical transformations}
\setcounter{equation}{0}

We now demonstrate that the addition of the boundary term (\ref{particle-boundary-term}) amounts to a canonical transformation.
Namely, the transformation of canonical variables
\bea
\left(\begin{matrix}
       p\\
       q
      \end{matrix}
\right) \mapsto
\left(\begin{matrix}
       P\\
       Q
      \end{matrix}
\right) :=
\left(\begin{matrix}
       p+S'_b(q)\\
       q\\
      \end{matrix}
\right),
\eea
is a canonical transformation since
\be
PdQ-pdq=dS_b(q)=dS_b(Q).
\ee
In particular, the symplectic form is preserved
\be
\Om=dp\wedge dq=dP\wedge dQ.
\ee
The generating function of this canonical transformation is
\be\label{particle-generator}
G(p,Q)=S_b(Q)+pQ,
\ee
so that
\be
P=\frac{\pa G}{\pa Q}, \quad q=\frac{\pa G}{\pa p}.
\ee
Since the generating function (\ref{particle-generator}) is time independent, this canonical transformation
does not change the Hamiltonian, which in the new coordinates, $(P,Q)$, becomes
\be
H(P,Q)=\frac12 (P-S'_b(Q))^2+ V(Q)=\frac12 P^2-PS'_b(Q)=\frac12 P^2+P\sqrt{-2V(Q)}.
\ee
Moreover, in terms of the new canonical variables we have
\be
(p+S'_b(q))dq-Hdt=pdq+dS_b(q)-Hdt=PdQ-Hdt,
\ee
and so the constraint (\ref{particle-renormalizability-condition}) can be written in the simplified form
\be\label{particle-renormalizability-condition-new}
P\dot{Q}-H\xrightarrow{t\to\infty} 0.
\ee

Although one can find a suitable time independent boundary term $S_b(q)$ that makes
the variational problem at infinity well defined by finding a suitable solution of the Hamilton-Jacobi equation, in some cases
it will be necessary to consider a boundary term that
depends explicitly both on $q$ and $t$. This is the case in asymptotically AdS gravity, for example, where
the explicit time dependence comes from the conformal anomaly \cite{Henningson:1998gx}. The reason behind the explicit
time dependence is not that one cannot find a time independent boundary term that makes the variational problem
well defined, but rather that a boundary term that simultaneously makes the variational problem well defined and
is local in boundary derivatives is necessarily explicitly time dependent. Since one in that context insists on
a local boundary term due to the holographic interpretation of the addition of such a boundary term, there is
no choice but to use a boundary term that explicitly depends on time.

Such more general boundary terms still amount to a canonical transformation, given by
\bea
\left(\begin{matrix}
       p\\
       q
      \end{matrix}
\right) \mapsto
\left(\begin{matrix}
       P\\
       Q
      \end{matrix}
\right) :=
\left(\begin{matrix}
       p+\frac{\pa S_b}{\pa q}\\
       q
      \end{matrix}
\right).
\eea
The new Hamiltonian, $K$, is not the same as the original one in this case though. The two Hamiltonians are related
by
\be
K=H-\frac{\pa G}{\pa t},
\ee
where
\be
G(p,Q,t)=S_b(Q,t)+pQ,
\ee
is the generating function of the canonical transformation. As before,
\be
P=\frac{\pa G}{\pa Q}, \quad q=\frac{\pa G}{\pa p}.
\ee

\subsection{Phase space and the space of asymptotic solutions}
\setcounter{equation}{0}

We have seen that the addition of the boundary term $S_b$ that makes the variational problem well defined amounts to
certain canonical transformation, but it is not clear yet what the significance of this canonical transformation is.

To address this question, we need to understand the relation between two symplectic manifolds, namely phase
space, $\cp$, parameterized by the symplectic coordinates $(p,q)$ on the one hand, and the space of asymptotic
solutions of the equation of motion as $t\to \infty$, $\cc$, which is parameterized by the integration constants
$(E,t_0)$, on the other. In the present case of a one dimensional point particle the system is integrable and the integration
constants $(E,t_0)$ actually parameterize exact solutions of the equation of motion, and not merely asymptotic ones,
but in general we only need to consider the space of asymptotic solutions.

We can now introduce a one-parameter family of symplectomorphisms
\bea
\varphi_t:\cc&\to& \cp\NO\\
(E,t_0)&\mapsto&(p,q)
\eea
given for every time $t$ by solving the equations of motion. More explicitly, this map is provided by the two
equations (\ref{time-integral}) and (\ref{particle-momentum}). Supplemented with the condition that $q(t)\to \infty$
as $t\to \infty$, which picks the plus sign in (\ref{particle-momentum}), this map is bijective for every fixed time $t$.
Moreover, the symplectic form $\Om_\cp$ on the cotangent bundle $\ct^*\cp$ induces via the (linear) pullback map
\bea
\varphi_t^*:\ct^*\cp&\to& \ct^*\cc\NO\\
\left(\begin{matrix}
         dp\\
         dq
        \end{matrix}
        \right)&\mapsto&
        \left(\begin{matrix}
         dE\\
         dt_0
        \end{matrix}
        \right),
\eea
a symplectic form, $\Om_\cc=\varphi_t^*\Om_\cp$, on the cotangent bundle $\ct^*\cc$. Namely,
\be
\varphi_t^*\Om_\cp=\varphi_t^*(dp\wedge dq)\stackrel{(\ref{particle-momentum})}{=} d\sqrt{2(E-V(q))}\wedge dq
= dE\wedge \frac{dq}{\sqrt{2(E-V(q))}}\stackrel{(\ref{time-integral})}{=} -dE\wedge dt_0 \equiv  \Om_\cc.
\ee

Let us now look at the form that the linear map $\vf^*_t$ takes asymptotically before and after the
canonical transformation induced by the boundary term $S_b$. From (\ref{particle-momentum}) and
(\ref{time-integral}) we easily see that for $t\to \infty$
\bea
&&dp\sim\frac{1}{\sqrt{-2V(q)}}dE+V'(q)dt_0,\NO\\
&&dq\sim -\sqrt{-2V(q)} dt_0,
\eea
and hence, before the canonical transformation,
\be
\vf^*_t\to \left(\begin{matrix}
                 \frac{1}{\sqrt{-2V(q)}} & V'(q) \\
                0 & -\sqrt{-2V(q)}
                \end{matrix}
\right).
\ee
After the canonical transformation, however,
\bea
&&dP=dp+S''_b(q)dq\sim\frac{1}{\sqrt{-2V(q)}}dE\NO\\
&&dQ\sim -\sqrt{-2V(q)} dt_0,
\eea
and so now
\be
\vf^*_t\to \left(\begin{matrix}
                 \frac{1}{\sqrt{-2V(q)}} & 0 \\
                0 & -\sqrt{-2V(q)}
                \end{matrix}
\right).
\ee
The pullback map between the cotangent bundle of phase space and the cotangent bundle of the space
of asymptotic solutions is now diagonal. We therefore see that the effect of the canonical transformation
implemented by the addition of the boundary term $S_b$ is to make the symplectic map $\varphi_t^*$ asymptotically
diagonal.

The conclusion of the above discussion is that the fact that the variational problem at infinity is not well defined,
the fact that the on-shell action is infinite and the fact that the symplectic map between the cotangent bundle of phase space
and the cotangent bundle of the space of asymptotic solutions of the equations of motion is not diagonal, are different manifestations of the same problem.  They are all simultaneously solved by a suitable canonical transformation that
corresponds to the addition of a certain boundary term to the action.

This can be understood as follows. In order to define the variational problem at infinity, with time independent
boundary conditions, we need to replace the phase space variables $(p,q)$ with some time independent variables
that yet capture the same dynamics. This requires that the variational problem is formulated as one over
variations that asymptotically approach generic solutions of the equations of motion. There is then a symplectic map
between phase space and the symplectic manifold of asymptotic solutions, which is parameterized in terms of time
independent variables. The variational problem can then be formulated in terms of these variables. However, in order
for boundary conditions imposed on the time independent variables parameterizing asymptotic solutions to be
expressible in covariant form purely in terms of the phase space variables $(p,q)$, and {\em not} involving explicitly time $t$,
it is essential that the symplectic map $\varphi_t^*$ be asymptotically diagonal.

Finally, let us emphasize that the requirement that the variational problem be well defined or that the symplectic map
$\varphi_t^*$ be asymptotically diagonal does not uniquely determine the canonical transformation or the boundary term
that needs to be added to the action. Another way to say this is that once a canonical transformation that renders the
variational problem well defined has been performed, there is still a subgroup of canonical transformations
that can be performed without spoiling the variational problem. These remaining canonical transformations can be classified
into two main types. One is rather trivial and is known in the literature as the `freedom of adding finite counterterms'.
This type of canonical transformations map the canonical coordinate $Q$ to itself, as the canonical transformation
required to make the variational problem well defined, and they originate in the fact that the boundary term to be added
is deduced by solving the Hamilton-Jacobi equation. Since this is a first order equation in $n$ variables, a generic solution
will contain $n$ arbitrary constants. This can be seen from the fact that Hamilton's principal function is the generator
of a canonical transformation that makes the canonical momenta constants. The arbitrariness in the solution
of the Hamilton-Jacobi equation is precisely the choice of these constants. In the example at hand this constant is related
to the energy $E$. We want the boundary term to be independent of the energy $E$, but any boundary term obtained as a solution
of the Hamilton-Jacobi equation with a {\em fixed} value of the energy $E$ is equally good, in that they all make the variational problem well defined. This type of ambiguity in the boundary term is lifted by requiring that the canonical transformation
corresponding to the boundary term strictly diagonalizes the symplectic map $\varphi_t^*$.

The second type of residual canonical transformations that leave the variational problem well defined involve both $P$ and $Q$
and correspond to different boundary conditions, such as Neumann or mixed boundary conditions. This class of canonical
transformations consists of arbitrary symplectic transformations in the space of asymptotic solutions. Such
boundary conditions are not generically allowed, however, unless both asymptotic solutions of the equations of motion
are normalizable. This is the case, for example, for massive scalar fields in AdS$_{d+1}$ with mass squared in the window
\cite{Breitenlohner:1982jf}
\be
-(d/2)^2\leq m^2 \leq -(d/2)^2+1,
\ee
where such more general boundary conditions are interpreted as multi-trace deformations of the dual QFT
\cite{Klebanov:1999tb, Witten:2001ua}, or for $U(1)$ gauge fields in AdS$_4$ where these boundary conditions are related to electric-magnetic duality \cite{Witten:2003ya}. The corresponding canonical transformation takes the form
\bea
\left(\begin{matrix}
       P\\
       Q
      \end{matrix}
\right) \mapsto
\left(\begin{matrix}
       \tilde{P}\\
       \tilde{Q}
      \end{matrix}
\right) :=
\left(\begin{matrix}
       Q\\
       -P-f'(Q)
      \end{matrix}
\right),
\eea
 where $f(Q)$ is an arbitrary function subject to the condition\footnote{This condition is actually slightly different in the
 case of scalar fields in AdS due to the contribution of the induced metric in the canonical momentum, but the essential
 idea is the same.} $f''(Q)\to 0$ as $Q\to\infty$. It is easy to check that under this canonical transformation, which corresponds
 to adding the boundary term
\be
 \tilde{S}_b=-PQ+f(Q)-Qf'(Q)
 \ee
 to the action, not only the map $\varphi_t^*$ remains diagonal as $t\to\infty$, but also the on-shell action remains finite. Note
 that for $f=0$, this canonical transformation corresponds to a Legendre transformation.

\subsection{The path integral and Schr\"odinger's equation}
\setcounter{equation}{0}

We conclude this section with a few remarks about how the above classical picture fits into the definition of a quantum
mechanical path integral, analogous to the path integral one is instructed to perform in the context of
the AdS/CFT correspondence in order to compute the generating functional of correlation functions in the dual QFT
\cite{Witten:1998qj}. Of course, no claim is made that such a path integral for the point particle computes anything in a holographically dual theory.

In particular, we want to evaluate the following path integral
\be\label{path-integral}
\lim_{t\to\infty}\ca(q,t;0,0)=\lim_{t\to\infty}\langle q|e^{-itH}|0\rangle =\lim_{t\to\infty}\int_0^q \cd q'e^{iS[q']},
\ee
where the integral is over paths that for large $t$ asymptote to classical solutions $q(t)$ of the equations of motion, with $q(t)\to\infty$ as $t\to\infty$. Note that the amplitude $\ca(q,t;0,0)$ is simply a wavefunction satisfying the Schr\"odinger equation. It can be expressed as
\be
\ca(q,t;0,0)=\int dEe^{-iEt}\langle q|E\rangle \langle E|0\rangle =\int dEe^{-iEt}\psi_E(q)\psi^*_E(0),
\ee
in terms of properly normalized energy eigenfunctions $\psi_E(q)$, which, for the class of potentials we specified above, are
$\d$-function normalizable. This can be seen by solving the time independent Schr\"odinger equation
\be
\left(-\frac12\frac{\pa^2}{\pa q^2}+V(q)\right)\psi_E(q)=E\psi_E(q),
\ee
in the WKB approximation to obtain
\be
\psi^{WKB}_E(q)=\frac{\rm const.}{(2(E-V(q)))^{1/4}}e^{i\int^q dq'\sqrt{2(E-V(q'))}}.
\ee
For $V(q)\to -\infty$ with $\int^qdq'1/\sqrt{-V(q')}\to \infty$ as $q\to\infty$, these WKB wavefunctions are $\d$-function normalizable, and hence so are the full wavefunctions $\psi_E(q)$.

However, the limit $t\to\infty$ of this amplitude as it stands is not well defined since the corresponding classical variational problem is not well defined for the action $S[q]$. Using the above $WKB$ wavefunctions we see that the amplitude in fact
behaves like
\be
\ca(q,t;0,0)\sim \psi^{WKB}_{E=0}(q(t))=\frac{\rm const.}{(-2V(q(t)))^{1/4}}e^{i\int^{q(t)} dq'\sqrt{-2V(q')}},
\ee
for large $t$, which is highly oscillatory.

Nevertheless, with the understanding of the variational problem we acquired in the previous subsection it is immediately clear how
to proceed in order to make the path integral (\ref{path-integral}) well defined. Firstly, we note that the amplitude
$\ca(q,t;0,0)$ can equivalently be written as
\be
\ca(q,t;0,0)=\int^{q_{cl}(t)} \cd p'\cd q'e^{i\int^t dt' (p'\dot{q}'-H(p',q'))},
\ee
which is valid for Hamiltonians even more general than (\ref{particle-Hamiltonian}). The results of the previous subsection
suggest that in order to impose the desired boundary conditions at $t\to\infty$ we must perform the path integral
over the canonical variables $(P,Q)$, which asymptotically diagonalize the symplectic map $\varphi_t^*$, and not over the
original $(p,q)$ variables. Note that even though the theory is of course invariant under canonical transformations, the
boundary conditions break this invariance. Boundary conditions on the path integral are imposed by going to the
corresponding canonical frame. Since the Jacobian of the canonical transformation from $(p,q)$ to $(P,Q)$ is $1$, we obtain
\be\label{path-integral-new}
\ca(Q,t;0,0)=\int^{Q_{cl}(t)}\cd P'\cd Q'e^{i\int^t dt' (P'\dot{Q}'-H(P',Q'))}.
\ee
This simply amounts to adding the boundary term $S_b$ to the original action, since
\be
\int^t dt' \left(P\dot{Q}-H(P,Q)\right)=\int^t dt' \left(p\dot{q}-H(p,q)\right)+ S_b(q(t)).
\ee
The canonical transformation and its relation to specifying boundary conditions, however, make the
addition of this boundary term natural and provide a deeper justification for it.

Again, the amplitude can be expressed as
\be
\ca(Q,t;0,0)=\int dE e^{-iEt}\Psi_E(Q)\Psi^*_E(0),
\ee
where $\Psi_E(Q)$ are now energy eigenfunctions satisfying the time independent Schr\"odinger equation
\be
\left(-\frac12\frac{\pa^2}{\pa Q^2}-2i\sqrt{-2V(Q)}\frac{\pa}{\pa Q}+\frac{iV'(Q)}{\sqrt{-2V(Q)}}\right)\Psi_E(Q)=E\Psi_E(Q),
\ee
in the $(P,Q)$ variables.
Solving this equation in the WKB approximation we now find
\be
\Psi^{WKB}_E(Q)=\frac{\rm const.}{(2(E-V(Q)))^{1/4}}e^{i\int^Q dQ'\left(\sqrt{2(E-V(Q'))}-\sqrt{-2V(Q')}\right)}.
\ee
Hence, for large $t$ the amplitude behaves as
\be
\ca(Q,t;0,0)\sim \Psi^{WKB}_{E=0}(Q(t))=\frac{\rm const.}{(-2V(Q(t)))^{1/4}},
\ee
which is well defined. Therefore, as anticipated, the effect of adding the boundary term $S_b$ in the classical
action and performing the path integral is the same as solving the Schr\"odinger equation in the correct variables,
$(P,Q)$, that asymptotically diagonalize the symplectic map $\f^*_t$.

\section{Holographic renormalization of diverse systems}
\label{examples}

The point particle example of the previous section demonstrates that the problems that holographic renormalization is
designed to address arise even in systems where there is no underlying holographic duality necessarily, and that our general
methodology for rendering the variational problem well defined applies to these cases as well. We would
now like to illustrate how the aspects of holographic renormalization discussed in the previous section apply to more
interesting and less simplistic examples, and especially in those cases where a holographic duality is expected to
play a role.

The aim of this section is to emphasize the diversity of systems to which holographic renormalization is applicable.
Our discussion of the various systems will therefore be rather brief, aimed mainly at demonstrating how the picture developed
in the previous section applies to each example. In particular, we will not be concerned with the computation of correlation
functions and the holographic dictionary, even in the cases where a holographic dual description is known or is expected to
exist. In the cases where such a holographic duality is known and understood, we will see that our general method
for regulating the variational problem uniquely singles out the correct variables that parameterize renormalized
observables in the dual theory. We expect the same to be true for systems to which our method is applicable, but
there is yet no known holographic dual.

We first discuss the case of a massive and massless scalar field in a fixed
AdS background, for which the boundary term required to make the variational problem well defined is already known. We will
then show that a {\em massive} scalar field in flat Euclidean space is not a very different problem and we can equally well apply
holographic renormalization to this case. The novel boundary term required to make the variational problem well defined
contains an infinite number of transverse derivatives, but it still admits a derivative expansion and preserves the symplectic form. We then proceed to consider a {\em massless} scalar field in flat Euclidean space,
which confronts us with new features. Namely, the asymptotic form of the solutions depends explicitly
on the Laplacian on the boundary sphere. In order to make the variational problem well defined in this case we will
need to reformulate the degrees of freedom in terms of an infinite set of symmetric traceless tensors, which are
essentially KK modes on the boundary sphere.  We will then briefly discuss asymptotically AdS gravity and the
role of diffeomorphism invariance, before concluding
with the application of holographic renormalization to the Polyakov action for closed strings. After showing
the role of holographic renormalization in the state-operator map in two dimensional CFTs, using the free boson
CFT example, we will consider closed strings in AdS, intersecting the boundary on an arbitrary curve. We will show
that our general method for regulating the variational problem is perfectly applicable, and the boundary term
required is uniquely determined to be the proper length of the intersection curve of the string with the AdS boundary.
This has direct application to the evaluation of Wilson loop expectation values \cite{Rey:1998bq, Drukker:1999zq} or gluon scattering amplitudes \cite{Alday:2007hr} in the dual $\cn=4$ super Yang-Mills theory. Further examples will appear in
future work \cite{Papadimitriou-AF,Papadimitriou-HQCD}.

\subsection{Scalar field in a fixed gravitational background}
\setcounter{equation}{0}

Let us first consider a self interacting scalar field in a fixed metric background. To avoid subtleties relating
to Lorentzian signature we will take the metric to be Euclidean here. We therefore consider the action
\be
S=\int d^{d+1}x\sqrt{g}\left(\frac12g^{\mu\nu}\pa_\m\f\pa_\nu \f+V(\f)\right),
\ee
where the metric takes the form
\be
ds^2=dr^2+\g_{ij}(r,x)dx^i dx^j,
\ee
$i,j=1,2,\ldots,d$, and the induced metric on the constant $r$ slices is given by
\be
\g_{ij}(r,x)=e^{2A(r)}\hat{g}_{ij}(x),
\ee
where $A(r)$ is some prescribed function of $r$. We will consider two particular cases, namely
\bea
AdS_{d+1}: & A(r)=r, & \hat{g}_{ij}(x)= \d_{ij},\NO\\
\mathbb{R}^{d+1}: & A(r)=\log r, & R[\hat{g}]_{ij}=(d-1)\hat{g}_{ij}.
\eea

The analysis of the variational problem is almost identical to the case of the point particle in the previous section, but with
some crucial differences relating to the fact that the boundary term required to make the variational problem well defined in this
case involves transverse derivatives. We start by reformulating the problem in Hamiltonian language in terms of the `time' $r$, with
the boundary located at $r\to\infty$. We therefore rewrite the action as
\be
S=\int^rdr'L=\int^rdr'd^dx\sqrt{\g}\left(\frac12\dot{\f}^2+\frac12\g^{ij}\pa_i\f\pa_j\f+V(\f)\right).
\ee
The canonical momentum conjugate to $\f$ then is
\be
\p=\frac{\d L}{\d \dot{\f}}=\sqrt{\g}\dot{\f}.
\ee
Adding a yet undetermined boundary term $S_b(\g,\f,r)$, which may depend explicitly on $r$ in addition to $\f$
and the induced metric $\g$, and considering a generic variation of the total action we get
\bea
\d (S+S_b)&=&-\int^r dr'd^dx\left(\frac{1}{\sqrt{\g}}\pa_r(\sqrt{\g}\dot{\f})+\square_\g\f-V'(\f)\right)\d\f\NO\\
&&+\left(L+\int d^dx\left(\dot{\f}\frac{\d S_b}{\d\f}+\dot{\g}_{ij}\frac{\d S_b}{\d\g_{ij}}\right)+\frac{\pa S_b}{\pa r}\right)\d r
+\int d^dx \left(\p+\frac{\d S_b}{\d\f}\right)\d\f.
\eea
The condition that determines the boundary term $S_b(\g,\f,t)$, therefore, is

\be
L+\int d^dx\left(\dot{\f}\frac{\d S_b}{\d\f}+\dot{\g}_{ij}\frac{\d S_b}{\d\g_{ij}}\right)+\frac{\pa S_b}{\pa r}=
\int d^dx \dot{\f}\left(\p+\frac{\d S_b}{\d\f}\right)-\left(H-\int d^dx\dot{\g}_{ij}\frac{\d S_b}{\d\g_{ij}}-\frac{\pa S_b}{\pa r}\right)\xrightarrow{r\to\infty}0,
\ee
or simply
\be\label{scalar-renormalizability-condition}
\frac{d}{dr}\left(S+S_b\right)\xrightarrow{r\to\infty}0.
\ee
The addition of the boundary term $S_b$, as in the point particle case, amounts to the phase space transformation
\bea\label{canonical-trans-scalar}
\left(\begin{matrix}
       \p\\
       \f
      \end{matrix}
\right) \mapsto
\left(\begin{matrix}
       \P\\
       \f
      \end{matrix}
\right):=
\left(\begin{matrix}
       \p+\frac{\d S_b}{\d\f}\\
       \f
      \end{matrix}
\right).
\eea
However, contrary to the point particle example, this transformation is not automatically canonical. It is a canonical
transformation if and only if it preserves the symplectic form, i.e.
\be
\Om=\int d^dx \d\p\wedge \d\f=\int d^dx \d\P\wedge \d\f.
\ee

Let us now turn to the problem of determining the boundary term that satisfies the condition
(\ref{scalar-renormalizability-condition}) and simultaneously preserves the symplectic from. Since we consider
variations within the space of scalar field configurations which asymptotically as $r\to\infty$ approach generic
solutions of the equation of motion, the action $S$ for large $r$ behaves as if it is evaluated on-shell on generic
asymptotic solutions and hence it can be replaced in the condition (\ref{scalar-renormalizability-condition}) with
Hamilton's principal function, $\cs$, which is a solution of the Hamilton-Jacobi equation. The Hamilton-Jacobi equation
for the present system can be derived from the relation
\be
\dot\cs=L=\int d^dx\left(\dot\f\frac{\d\cs}{\d\f}+\dot\g_{ij}\frac{\d\cs}{\d\g_{ij}}\right),
\ee
where we have taken $\cs$ to have no explicit $r$ dependence since it corresponds to
Hamilton's principal function for a Lagrangian that is diffeomorphism covariant. Writing
\be
\p=\sqrt{\g}\dot\f=\frac{\d\cs}{\d\f},
\ee
this equation becomes
\be
\int d^dx\left[\sqrt{\g}\left(\frac12\left(\frac{1}{\sqrt{\g}}\frac{\d \cs}{\d\f}\right)^2-\frac12\g^{ij}\pa_i\f\pa_j\f-V(\f)\right)+2\dot{A}\g_{ij}\frac{\d\cs}{\d\g_{ij}}\right]=0.
\ee
This is the Hamilton-Jacobi equation for the scalar field in the fixed gravitational background, which can be rewritten in
the more useful form
\be\label{HJ-fixed-background}
\sqrt{\g}\left(\frac12\left(\frac{1}{\sqrt{\g}}\frac{\d \cs}{\d\f}\right)^2-\frac12\g^{ij}\pa_i\f\pa_j\f-V(\f)\right)+2\dot{A}\d_\g \cl=\pa_i v^i,
\ee
where
\be
\cs=\int d^dx \cl,
\ee
and
\be
\d_\g=\int d^dx \g_{ij}\frac{\d}{\d\g_{ij}}.
\ee
The term $\pa_iv^i$ on the RHS is a total derivative that can be arbitrary, but which generically needs to be taken into account
when trying to solve (\ref{HJ-fixed-background}). It is not difficult to solve this equation iteratively, for example in a
derivative expansion, for a general potential $V(\f)$.\footnote{We refer to \cite{Papadimitriou-HQCD} for the closely
related problem of a scalar with a generic potential coupled to dynamical gravity.} However, since tackling a general potential would not add anything essential to the present exposition, we will consider the simple, yet far from trivial, case of
a free scalar field with the potential
\be
V(\f)=\frac12 m^2\f^2.
\ee
The great simplification that results from this potential is that we can solve the corresponding Hamilton-Jacobi
equation exactly, to all orders in transverse derivatives.

\subsubsection{Massive and massless scalar in AdS$_{d+1}$ background}

The equation of motion of a free massive scalar field on a fixed AdS$_{d+1}$ background is
\be\label{eom-ads-scalar}
\ddot\f+d \dot\f+e^{-2r}\square_\d \f-m^2\f=0,
\ee
where $\square_\d$ is the Laplacian on the flat transverse space. Our first task is to construct the space of
asymptotic solutions of this equation, $\cc$, and to evaluate the pullback of the symplectic form on this space.  The most general
asymptotic solution of (\ref{eom-ads-scalar}) takes the form \cite{Skenderis:2002wp}
\be\label{exp-ads-scalar}
\f(r,x)=e^{-(d-\D)r}\left(\f\sub{0}(x)+e^{-2r}\f\sub{2}(x)+\cdots+e^{-(2\D-d)r}\left(-2 r\psi\sub{2\D-d}(x)
+\f\sub{2\D-d}(x)\right)+\cdots\right),
\ee
where $m^2=\D(\D-d)$ and we have taken $d/2<\D<d$. This asymptotic expansion takes a slightly different form for the
cases $\D=d/2$ and $\D=d$, but the analysis is essentially the same. Inserting this formal asymptotic expansion into the
equation of motion one finds that the functions $\f\sub{0}(x)$ and $\f\sub{2\D-d}(x)$ are arbitrary, while the functions
$\f\sub{n}(x)$, $0<n<2\D-d$, and $\psi\sub{2\D-d}(x)$ are uniquely determined locally in terms of $\f\sub{0}(x)$. The
functions $\f\sub{n}(x)$, with $n>2\D-d$ are also determined uniquely in terms of both $\f\sub{0}(x)$ and $\f\sub{2\D-d}(x)$.
The space of asymptotic solutions, $\cc$, is therefore parameterized by the arbitrary functions $\f\sub{0}(x)$ and
$\f\sub{2\D-d}(x)$.

The asymptotic expansion (\ref{exp-ads-scalar}), complemented with the expression $\p=\sqrt{\g}\dot\f$ for the canonical momentum
conjugate to $\f$, provides a map $\vf_r: \cc\to\cp$ from the space of asymptotic solutions to the phase space. The pullback
map $\vf_r^*:\wedge^n\ct^*\cp\to \wedge^n\ct^*\cc$ then maps the symplectic form $\Om_\cp$ on phase space to a symplectic form on
the space of asymptotic solutions. Namely,
\be
\vf^*_r\Om_\cp=\vf^*_r\int d^dx \d\p\wedge \d\f=(d-2\D)\int d^dx \d\f\sub{2\D-d}\wedge \d\f\sub{0},
\ee
where we have used the fact that the terms $\f\sub{n}(x)$, $0<n<2\D-d$, and
$\psi\sub{2\D-d}(x)$ are locally expressed in terms of $\f\sub{0}(x)$, and that the symplectic
form $\Om_\cp$ is independent of the radial coordinate $r$ \cite{Lee:1990nz,Papadimitriou:2005ii}, and hence the pullback
of the symplectic form can be evaluated by taking the limit $r\to\infty$.

We turn next to the task of determining the boundary term $S_b$. The Hamilton-Jacobi equation (\ref{HJ-fixed-background}) in this case becomes
\be\label{HJ-fixed-AdS}
\sqrt{\g}\left(\frac12\left(\frac{1}{\sqrt{\g}}\frac{\d \cs}{\d\f}\right)^2-\frac12\g^{ij}\pa_i\f\pa_j\f
-\frac12 m^2\f^2\right)+2\d_\g \cl=\pa_i v^i.
\ee
Inserting an ansatz of the form
\be\label{HJ-ansatz}
\cs=\frac12\int d^dx \sqrt{\g}\f f(-\square_\g)\f,
\ee
we find that it solves the Hamilton-Jacobi equation provided the function $f(x)$ satisfies \cite{Papadimitriou:2003is}
\be
f^2(x)+d f(x)-m^2-x-2xf'(x)=0.
\ee
The general solution of this equation is
\be\label{sol-scalar-AdS}
f(x)=-\frac d2-\frac{\sqrt{x} \left(K_k'(\sqrt{x})+c I_k'(\sqrt{x})\right)}{K_{k}(\sqrt{x})+c I_k(\sqrt{x})},
\ee
where $k=\D-d/2>0$, $c$ is an arbitrary constant, and $I_k(x)$ and $K_k(x)$ denote the modified Bessel function of
the first and second kind respectively. Using the asymptotic behaviors as $x\to 0$
\be
K_0(x)\sim -\log x,\quad K_k(x)\sim \frac{\G(k)}{2}\left(\frac x2\right)^{-k},\quad k>0,\quad
\quad I_k(x)\sim \frac{1}{\G(k+1)}\left(\frac x2\right)^k,
\ee
we see that $K_k(x)$ dominates in $f(x)$ as $x\to 0$, unless $|c|\to \infty$. In particular, we find
\be
f(x)\stackrel{x\to 0}{\sim} \left\{\begin{matrix}
          -\frac d2+k=-(d-\D),\quad |c|<\infty,\\
          -\frac d2-k=-\D,\quad |c|\to\infty.
                \end{matrix}
\right.
\ee
Since,
\be
\dot{\f}=\frac{1}{\sqrt{\g}}\frac{\d\cs}{\d\f},
\ee
we see that the two asymptotic solutions for $f(x)$ correspond to $\f\sim e^{-(d-\D)r}$ and $\f\sim e^{-\D r}$ respectively,
which are precisely the asymptotic behaviors of the two linearly independent solutions of the equation of motion.
The solution for $f(x)$ with $|c|<\infty$ corresponds to the asymptotically dominant mode. Hence, in order to make the
variational problem well defined for generic solutions of the equation of motion we have no choice but demand that $|c|<\infty$.

Expanding the solution for $f(x)$ with $|c|<\infty$ for small $x$ and taking $k$ to be an integer we obtain,
\bea
f(x)&=&-(d-\D)+\frac{x}{(2\D-d-2)}-\frac{x^2}{(2\D-d-2)(2\D-d-4)}+\cdots+\frac{(-1)^k}{2^{2k-1}\G(k)^2}x^{k}\log x\NO\\
&&+ \left(a(k)-\frac{c}{2^{2k-2}\G(k)^2}\right)x^k+\cdots,
\eea
where $a(k)$ is a known function of $k$, whose explicit form we will not need, and the dots denote asymptotically
subleading terms. A number of comments are in order here. Firstly, this solution depends explicitly on the undetermined
constant $|c|<\infty$. Secondly, this solution seems to lead to a non-local boundary term due to the logarithmic term.
And finally, one may worry that higher terms in this asymptotic expansion need to be considered. Fortunately, all these
issues can be addressed by noticing that the contribution of the last term to the boundary term is proportional to
\be
\int d^dx \sqrt{\g}\f(-\square_\g)^k\f,
\ee
which, taking into account the asymptotic behavior of the scalar and of the induced metric, can be easily seen to have
a finite limit as $r\to\infty$. Such terms, therefore, correspond to adding finite local contributions to the boundary term $S_b$.
We conclude that higher order terms in the asymptotic expansion of $f(x)$ need not be considered since they would
give rise to a vanishing contribution to $S_b$ in the limit $r\to\infty$. Moreover, the arbitrariness in the value of $c$
is not a problem because different values of $c$ lead to boundary terms $S_b$ which differ by a finite local term. Any
value of $|c|<\infty$, therefore, is equally acceptable since the corresponding boundary term makes the variational problem
well defined. Finally, coming to the apparent non-locality of the boundary term we have deduced above, we notice
that the logarithmic term can be written as
\be
(-\square_\g)^k\log(-\square_\g)=(-\square_\g)^k\left(\log(\m^2e^{-2r})+\log(-\square_{\d}/\m^2)\right),
\ee
where $\m^2$ is an arbitrary scale and $\square_{\d}=\pa_i\pa_i$ denotes the Laplacian in the flat transverse space.
Crucially, the non-local part gives rise to a finite contribution in Hamilton's principal function and so it
can be omitted from the boundary term $S_b$. The most general local boundary term that makes the variational problem well
defined is therefore  \cite{Skenderis:2002wp,Papadimitriou:2003is}
\bea\label{ads-scalar-boundary-term}
S_b(\g,\f,r)&=&-\frac12\int d^dx\sqrt{\g}\f\left(-(d-\D)+\frac{-\square_\g}{(2\D-d-2)}-\frac{(-\square_\g)^2}{(2\D-d-2)(2\D-d-4)}+\cdots\right.\NO\\
&&\left.+\frac{(-1)^k}{2^{2k-1}\G(k)^2}(-\square_\g)^{k}\log(\m^2e^{-2r})+\x (-\square_\g)^{k}\right)\f,
\eea
where we have allowed for a finite boundary term with arbitrary coefficient $\x$. By construction, this boundary term
is the most general local boundary term that satisfies the condition (\ref{scalar-renormalizability-condition}). Notice
that although it is possible to find a boundary term that simultaneously makes the variational problem well defined
and is also local in transverse derivatives, this is only at the cost of introducing explicit dependence in the radial
coordinate, $r$. This is precisely the origin of the holographic conformal anomaly \cite{Henningson:1998gx}.

We can now check explicitly that the transformation (\ref{canonical-trans-scalar}) induced by the boundary term $S_b$
is a canonical transformation. This follows from the fact that,
\be
\int d^dx \d\P\wedge \d\f=\int d^dx \left(\d\p+\frac{\d^2 S_b}{\d\f^2}\d\f\right)\wedge \d\f,
\ee
where
\bea
&&\int d^dx \left(\frac{\d^2 S_b}{\d\f^2}\d\f\right)\wedge \d\f= \int d^dx \d_1\left(\frac{\d S_b}{\d\f}\right)\d_2\f-(1\leftrightarrow 2)\NO\\
&&=\d_1\int d^dx \frac{\d S_b}{\d\f}\d_2\f-(1\leftrightarrow 2)=(\d_1\d_2-1\leftrightarrow 2)S_b=0.
\eea
Hence, the symplectic form is preserved under the transformation (\ref{canonical-trans-scalar}), which is
therefore canonical. We should emphasize at this point that the fact that the symplectic form is invariant
is not a consequence of $S_b$ being local in transverse derivatives. This can be demonstrated by considering
a generic non-local boundary term of the form
\be
S_b=\frac12\int_{\S_r} d^dx d^dy \f(r,x)K(|x-y|;r)\f(r,y),
\ee
where $K(|x-y|;r)$ is some generic kernel, and repeating the calculation we just did. We again find that
the transformation corresponding to such a boundary term still preserves the symplectic from, and hence
it is a canonical transformation. Of course, in the context of the AdS/CFT correspondence, adding
such a non local boundary term will completely change the dual field theory. From the point of view
of making the bulk variational problem well defined, however, such non local boundary terms are allowed as long
as they correspond to canonical transformations. In particular, in the present case we could have taken the boundary
term to be (\ref{HJ-ansatz}), with the function $f$ being given by (\ref{sol-scalar-AdS}), without truncating
the asymptotic expansion of $f$ as we did above. Such a boundary term contains a non-local contribution of the form
$\f(-\square_\g)^k\log(-\square_\g)\f$, but, as we have just argued, it still preserves the symplectic form.
The choice in the boundary term $S_b$ that makes the variational problem well defined simply means that
there is a choice of acceptable boundary conditions, corresponding to the different boundary terms. In
the present case, the AdS/CFT dictionary singles out the local boundary term, but still allows for the
scheme dependence corresponding to the choice of the constant $\x$. As we will see next, the
requirement that the map $\vf_r^*$ be diagonal, not only singles out the local boundary term, but
also uniquely fixes the value of $\x$.

Given that $S_b$ induces a canonical transformation, it is legitimate to ask what is the effect of
this canonical transformation on the map $\vf_r^*:\ct^*\cp\to \ct^*\cc$. Using the asymptotic expansion (\ref{exp-ads-scalar})
and the relation $\p=\sqrt{\g}\dot\f$ we find
\be
\vf^*_r\left(\begin{matrix}
              \d \p\\
              \d\f
             \end{matrix}
\right)=
\left(\begin{matrix}
       -(d-\D)e^{\D r}\d\f\sub{0}(x)+\cdots+e^{(d-\D)r}\left(-2 \d\psi\sub{2\D-d}(x)
-\D\d\f\sub{2\D-d}(x)\right)+\cdots \\
e^{-(d-\D)r}\d\f\sub{0}(x)+\cdots
      \end{matrix}
\right),
\ee
where $\d\psi\sub{2\D-d}(x)\propto \square_\d^k\d\f\sub{0}(x)$ and the dots denote asymptotically subleading terms.
However, after the canonical transformation we obtain instead
\be
\vf^*_r\left(\begin{matrix}
              \d \P\\
              \d\f
             \end{matrix}
\right)=
\left(\begin{matrix}
       (d-2\D)e^{(d-\D)r}\d\f\sub{2\D-d}(x)+\l(k,\x) e^{(d-\D)r}\square_\d^k\d\f\sub{0}(x)+\cdots \\
e^{-(d-\D)r}\d\f\sub{0}(x)+\cdots
      \end{matrix}
\right),
\ee
where $\l(k,\x)$ is a function of $k=\D-d/2$ and of the constant $\x$ in the boundary term
(\ref{ads-scalar-boundary-term}) which we need not compute explicitly here. It suffices to note that by a choice
of $\x$ in (\ref{ads-scalar-boundary-term}) we can set $\l(k,\x)=0$. Modulo the freedom of adding a finite
local boundary term, therefore, the effect of the canonical transformation induced by $S_b$ is to diagonalize
the symplectic map $\vf_r^*:\ct^*\cp\to \ct^*\cc$. We therefore see in this case too that requiring that this
map be diagonal is equivalent to requiring that the variational problem is well defined. The difference with
respect to the point particle example of the previous section is that here the requirement of $\vf_r^*$ to
be asymptotically diagonal is somewhat stronger than the requirement that the variational problem be well defined,
in that the former uniquely fixes the ``scheme dependence'', i.e. the freedom in the choice of $\x$ in
the boundary term (\ref{ads-scalar-boundary-term}), left by the latter.

\subsubsection{Massive scalar in $\mathbb{R}^{d+1}$ background}

Let us now repeat the above analysis for a massive scalar in a fixed $\mathbb{R}^{d+1}$ background. The
equation of motion for such a scalar is
\be\label{eom-flat-scalar}
\ddot\f+\frac dr\dot\f +\frac{1}{r^2}\square_{\hat g}\f-m^2\f=0,
\ee
where $\hat g_{ij}$ is the metric on $S^{d}$. Our first task again is to construct the space of
asymptotic solutions, $\cc$, and evaluate the pull back of the symplectic form on this space.

Taking $m>0$, the most general asymptotic solution of (\ref{eom-flat-scalar}) takes the form
\bea\label{exp-flat-scalar-m}
\f(r,x)&=&r^{-d/2}e^{mr}\left(1-\frac{1}{2mr}\left(\left(\frac{d-1}{2}\right)^2-\square_{\hat g}\right)+\co(1/r^2)\right)\f_+(x)
\NO\\
&&+r^{-d/2}e^{-mr}\left(1+\frac{1}{2mr}\left(\left(\frac{d-1}{2}\right)^2-\square_{\hat g}\right)+\co(1/r^2)\right)\f_-(x),
\eea
where $\f_\pm(x)$ are arbitrary functions on $S^d$. This asymptotic expansion can be easily obtained to any desired order by expanding asymptotically the general solution
\be
\f(r,x)=r^{-(d-1)/2}\left(I_\n(mr)\f_+(x)+K_\n(mr)\f_-(x)\right),
\ee
of (\ref{eom-flat-scalar}), where $\n=\sqrt{(d-1)^2/4-\square_{\hat g}}$. Note that only $\n^2$ appears in
this expansion to any order. This expansion again, complemented with the expression $\p=\sqrt{\g}\dot\f$ for the canonical
momentum conjugate to $\f$, provides a map $\vf_r: \cc\to\cp$ from the space of asymptotic solutions to phase space. The pullback
of the symplectic form then gives
\be
\vf^*_r\Om_\cp=\vf^*_r\int d^dx \d\p\wedge \d\f=2m\int d^dx \d\f_+\wedge \d\f_-,
\ee
where again we have used the fact that the symplectic form is independent of $r$ and so can be evaluated in the limit $r\to\infty$.

Having constructed the symplectic form on the space of asymptotic solutions, we now determine the appropriate boundary term
that makes the variational problem  well defined. Since
\be
R[\g]=e^{-2A}R[\hat{g}]=e^{-2A}d(d-1),
\ee
with $A(r)=\log r$, the Hamilton-Jacobi equation (\ref{HJ-fixed-background}) becomes
\be\label{HJ-fixed-flat}
\sqrt{\g}\left(\frac12\left(\frac{1}{\sqrt{\g}}\frac{\d \cs}{\d\f}\right)^2-\frac12\g^{ij}\pa_i\f\pa_j\f-\frac12 m^2\f^2\right)
+2\sqrt{\frac{R[\g]}{d(d-1)}}\d_\g \cl=\pa_i v^i.
\ee
We look again for a solution of the Hamilton-Jacobi equation of the form
\be\label{flat-ansatz}
\cs=\frac12\int d^dx \sqrt{\g}\f f(-\square_\g,R[\g])\f.
\ee
Inserting this ansatz into (\ref{HJ-fixed-flat}) we find that the function $f(x,y)$ must satisfy
\be
f^2-x-m^2+\sqrt{\frac{y}{d(d-1)}}\left(d f-2xf_x-2yf_y\right)=0.
\ee
Substituting,
\be
f(x,y)=\sqrt{\frac{y}{d(d-1)}} g(x,y),
\ee
in this equation, it becomes
\be
g^2+(d-1)g-2xg_x-2yg_y-\frac{d(d-1)}{y}(x+m^2)=0.
\ee
Changing variables to $z=d(d-1)x/y$ and $w=d(d-1)/y$ we then arrive at the first order ODE
\be
g^2+(d-1)g+2 w\pa_w g-wm^2-z=0.
\ee
With $m>0$ again, the general solution  is
\be
g=-\frac{d-1}{2}+\frac{m\sqrt{w}\left(I'_\n(m\sqrt{w})+c K'_\n(m\sqrt{w})\right)}{I_\n(m\sqrt{w})+c K_\n(m\sqrt{w})},
\ee
where $c$ is an arbitrary constant and
\be
\n=\sqrt{\left(\frac{d-1}{2}\right)^2+z}.
\ee
Hence,
\be\label{flat-sol}
f=-\frac{d-1}{2\sqrt{w}}+\frac{m\left(I'_\n(m\sqrt{w})+c K'_\n(m\sqrt{w})\right)}{I_\n(m\sqrt{w})+c K_\n(m\sqrt{w})}.
\ee
Using the asymptotic behavior of the modified Bessel functions for large argument \cite{abramowitz+stegun},
\bea
&&I_\n(m\sqrt{w})\sim \frac{e^{m\sqrt{w}}}{\sqrt{2\p m\sqrt{w}}}\left(1-\frac{4\n^2}{8m\sqrt{w}}+\co(1/(m\sqrt{w})^2)\right),\NO\\
&&K_\n(m\sqrt{w})\sim \sqrt{\frac{\p}{2 m\sqrt{w}}}e^{-m\sqrt{w}}\left(1+\frac{4\n^2}{8m\sqrt{w}}+\co(1/(m\sqrt{w})^2)\right),
\eea
we find that as $w\to\infty$, which corresponds to $r\to\infty$,
\be
f=\left\{\begin{matrix}
       m-\frac{d}{2\sqrt{w}}+\co(1/w), & |c|<\infty,\\
       -m-\frac{d}{2\sqrt{w}}+\co(1/w), & |c|\to\infty.
         \end{matrix}
\right.
\ee
Since
\be
\dot{\f}=\frac{1}{\sqrt{\g}}\frac{\d\cs}{\d\f},
\ee
and $w=d(d-1)/R[\g]=e^{2A}=r^2$, we conclude that these two asymptotic behaviors of $f$ correspond respectively to
$\f\sim r^{-d/2}e^{\pm mr}$. These are precisely the two linearly independent modes of the equation of motion. We
have taken $m>0$ and so the solution for $f$ with $|c|<\infty$ corresponds to the asymptotically leading mode. It
follows that in order to make the variational problem well defined for generic solutions of the equations of
motion we must take $|c|<\infty$ for the boundary term. As for the AdS case, however, the precise value of $c$ is
not fixed by the requirement of making the variational problem well defined. Indeed, any finite value of $c$ will do.
To see this we again note that for large $w$
\be
f\sim-\frac{d-1}{2\sqrt{w}}+\frac{m I'_\n(m\sqrt{w})}{I_\n(m\sqrt{w})}-2\p m c e^{-2m\sqrt{w}}.
\ee
Inserting the term proportional to $c$ into Hamilton's principal function we see that it contributes a term
\be
\cs_c=-mc\p\int d^dx \sqrt{\g}\f^2 e^{-2m \sqrt{d(d-1)/R[\g]}},
\ee
which, using the asymptotic behavior of the induced metric and of the scalar field deduced above, we see that
it is finite as $r\to\infty$. We therefore confirm that $c$ corresponds to the generic non-uniqueness of
the boundary term required to make the variational problem well defined, and which is analogous to
what happens in AdS.

The outcome of the above discussion is that for $m>0$ we can take the boundary term that makes
the variational problem well defined to be given by (\ref{flat-ansatz}) with $f$ as in
(\ref{flat-sol}) with $c=0$. Any value of $|c|<\infty$ is equally good for making the variational problem well defined,
but the value $c=0$, as we shall see, is the unique value that diagonalizes the map $\vf_t^*$. The
boundary term that makes the variational problem for a {\em massive} scalar field in $\mathbb{R}^{d+1}$ well defined is
\be\label{flat-boundary-term}
S_b=-\frac12\int d^dx\sqrt{\g}\f\left(-\sqrt{\frac{(d-1)R[\g]}{4d}}+\frac{mI'_{\n}\left(m\sqrt{\frac{d(d-1)}{R[\g]}}\right)}{I_\n\left(m\sqrt{\frac{d(d-1)}{R[\g]}}\right)}\right)\f,
\ee
where
\be
\n=\sqrt{\left(\frac{d-1}{2}\right)^2-\frac{d(d-1)}{R[\g]}\square_\g}.
\ee
The crucial difference with the AdS case is that the boundary term required to make the variational problem
well defined necessarily contains an infinite number of transverse derivatives. Nevertheless, it still admits a
derivative expansion, namely
\be
S_b=-\frac12\int d^dx\sqrt{\g}\f\left(m-\sqrt{\frac{dR[\g]}{4(d-1)}}+\frac{(d-2)R[\g]}{8(d-1)m}-\frac{1}{2m}\square_\g
+\co\left(\frac{R^{3/2}}{m^2},\frac{R^{1/2}(-\square_\g)}{m^2}\right)\right)\f,
\ee
and it can be checked that this boundary term preserves the symplectic form and hence corresponds to
a canonical transformation.

Finally, applying the map $\vf_r^*:\ct^*\cp\to \ct^*\cc$ after the canonical transformation corresponding to
$S_b$ in (\ref{flat-boundary-term}) we get
\be
\vf^*_r\left(\begin{matrix}
              \d \P\\
              \d\f
             \end{matrix}
\right)=
\left(\begin{matrix}
       -2m r^{d/2}e^{-mr}\d\f_-(x)+\cdots \\
       r^{-d/2}e^{mr}\d\f_+(x)+\cdots
      \end{matrix}
\right).
\ee
Hence, once more, we see that with the choice $c=0$ for the boundary term, the corresponding canonical transformation
diagonalizes the symplectic map $\vf_r^*$.

\subsubsection{Massless scalar in $\mathbb{R}^{d+1}$ background}

We have so far seen that the massive scalar field in $\mathbb{R}^{d+1}$ is not so much different from
a scalar field in AdS, except from the fact that the boundary term required to make the variational problem
well defined
necessarily contains an infinite number of transverse derivatives. We now come to address the case of
a massless scalar field in $\mathbb{R}^{d+1}$, which, as will see, poses some drastic qualitative differences which
have to be understood. The issues that arise in this case are similar to those arising when considering the variational
problem of Type IIB supergravity in asymptotically $AdS_5\times S^5$ backgrounds. Our approach in tackling them here for
the massless scalar field will be conceptually similar to the
analysis in \cite{deBoer:2003vf}, but with some crucial differences.

Setting $m=0$ in (\ref{eom-flat-scalar}) we immediately see that the general solution of the equation of motion takes
the from
\be\label{sol-flat-scalar-m=0}
\f(r,x)=r^{-(d-1)/2}\sum_{\n}\left(r^{\n}\f_+^{(\n)}(x)+r^{-\n}\f_-^{(\n)}(x)\right),
\ee
where $\f_\pm^{(\n)}(x)$ satisfy
\be\label{flat-scalar-constraint}
\square_{\hat g} \f_\pm^{(\n)}(x) =\left(\left(\frac{d-1}{2}\right)^2-\n^2\right)\f_\pm^{(\n)}(x).
\ee
In other words, $\f_\pm^{(\n)}(x)$ can be written as a generic linear combination of spherical harmonics on $S^d$ with
a fixed eigenvalue of the Laplacian. The dramatic new feature of the solution  (\ref{sol-flat-scalar-m=0}) is that
the asymptotic behavior depends on the eigenvalue of the Laplacian in the transverse space. The fundamental
problem with this behavior is that the map $\vf^*_r$ corresponding to the asymptotic solution
(\ref{sol-flat-scalar-m=0}) does {\em not} induce a well defined symplectic form on the space of asymptotic solutions,
as can be checked explicitly. Before we proceed with trying to make the variational problem well defined, therefore,
we must understand how we can construct a well defined space of asymptotic solutions with a non degenerate
symplectic form.

In order to achieve this we must express the scalar field in terms of degrees of freedom on $S^d$ that trivialize the
constraint  (\ref{flat-scalar-constraint}). In other words, the scalar field should be expressed in terms
of a set of degrees of freedom which are eigenfunctions of the Laplace operator on $S^d$. There is a unique solution to this
problem, which is to expand the scalar field in terms of harmonic polynomials on $S^d$. Namely, introducing coordinates
$\{y^a\}\in \mathbb{R}^{d+1}$, $a=1,\ldots,d+1$, the space of Harmonic polynomials on $\mathbb{R}^{d+1}$ of degree $\ell$,
$\ch_{\ell}(\mathbb{R}^{d+1})$, consists of homogeneous polynomials of degree $\ell$ of the form
\be
\wp^{(\ell)}(y)=\frac{1}{\ell !}\wp^{(\ell)}_{a_1a_2\ldots a_\ell}y^{a_1}y^{a_2}\cdots y^{a_\ell},
\ee
where $\wp^{(\ell)}_{a_1a_2\ldots a_\ell}$ is a totally symmetric and traceless tensor. The tracelessness
of this tensor ensures that $\wp^{(\ell)}(y)$ is harmonic with respect to the Laplacian
$\square_{\mathbb{R}^{d+1}}=\pa_{y^a}\pa_{y^a}$. Now, writing $y^a=\r x^a$, where $x^a$ satisfy the constraint
$x^ax^a=1$, we can write the Laplacian on $\mathbb{R}^{d+1}$ as
\be
\square_{\mathbb{R}^{d+1}}=\r^{-d}\pa_\r(\r^d\pa_\r)+\r^{-2}\square_{S^d}.
\ee
Applying this on the harmonic polynomial $\wp^{(\ell)}(y)$ and then pulling back the resulting expression on $S^d$ we find
that
\be
\square_{S^d}\wp^{(\ell)}(x)=-\ell(\ell+d-1)\wp^{(\ell)}(x).
\ee
This result is precisely what we were looking for. The degrees of freedom that trivialize the constraint
(\ref{flat-scalar-constraint}) are harmonic polynomials on $\mathbb{R}^{d+1}$ pulled back on $S^d$. We can therefore
expand the scalar field as
\be
\f(r,x)=\sum_{\ell=0}^\infty \wp^{(\ell)}(r,x)=\sum_{\ell=0}^\infty
\frac{1}{\ell !}\wp^{(\ell)}_{a_1a_2\ldots a_\ell}(r)x^{a_1}x^{a_2}\cdots x^{a_\ell},
\ee
where each polynomial $\wp^{(\ell)}(r,x)$ satisfies the constraint (\ref{flat-scalar-constraint}) with $\n=(d-1)/2+\ell$.
The only other ingredient we need is to evaluate the moment integral
\be
\ci_{a_1a_2\ldots a_{2k}}\equiv \int_{S^d} \m(x) x^{a_1}x^{a_2}\cdots x^{a_{2k}},
\ee
where $\m(x)$ is the appropriate measure on $S^d$, defined such that
\be
\int_{S^d} \m(x)=\frac{2\p^{(d+1)/2}}{\G\left(\frac{d+1}{2}\right)}.
\ee
Note that the integral of an odd number of $x^a$s vanishes. Now, $\ci_{a_1a_2\ldots a_{2k}}$ must be a
totally symmetric Cartesian tensor, and hence
\be
\ci_{a_1a_2\ldots a_{2k}}=c_k(d)\sum_{\s\in B_{2k}}\d^{a_{\s(1)}a_{\s(2)}}\d^{a_{\s(3)}a_{\s(4)}}
\cdots \d^{a_{\s(2k-1)}a_{\s(2k)}},
\ee
where $B_{2k}$ denotes the $(2k-1)!!$ dimensional conjugacy class of the symmetric group $S_{2k}$, corresponding
to the partition $2k=2+2+\cdots+2$. Since the $x^a$s satisfy the constraint $x^ax^a=1$, contracting say the last two
indices in $\ci_{a_1a_2\ldots a_{2k}}$ must give identically $\ci_{a_1a_2\ldots a_{2k-2}}$. This implies that the
coefficient $c_k$ satisfies the recursion relation
\be
c_k(d)=\frac{c_{k-1}(d)}{d+1+2k-2},
\ee
which can be solved to obtain
\be
c_k(d)=\frac{2\p^{(d+1)/2}}{\G\left(\frac{d+1}{2}\right)}
\frac{2^{k-1}\G\left(\frac d2 +k\right)\G(d+1)}{\G\left(\frac d2+1\right)\G(d+2k)}=\frac{2\p^{(d+1)/2}}{2^k\G\left(\frac{d+1}{2}+k\right)}.
\ee
Using this result now we find that for two harmonic polynomials $\wp^{(\ell)}(x)$ and $\Im^{(\ell')}(x)$,
\be
\int_{S^d} \m(x) \wp^{(\ell)}(x)\Im^{(\ell')}(x)=\frac{2\p^{(d+1)/2}}{2^\ell\G\left(\frac{d+1}{2}+\ell\right)}
\frac{\d_{\ell\ell'}}{\ell !}\wp^{(\ell)}_{a_1a_2\ldots a_\ell}\Im^{(\ell)a_1a_2\ldots a_\ell}
\equiv \d_{\ell\ell'} \wp^{(\ell)}\cdot\Im^{(\ell)}.
\ee

We now have the necessary ingredients to reformulate the original problem in terms of the new degrees of freedom. Inserting
the decomposition of the scalar field in terms of the symmetric traceless tensors into the scalar Lagrangian and using
the above orthonormality property of the harmonic polynomials we obtain the action
\be
S=\sum_{\ell=0}^\infty \int dr e^{dA(r)}\left(\frac12\dot{\wp}^{(\ell)}\cdot\dot{\wp}^{(\ell)}+
\frac12 e^{-2A(r)} \ell(\ell+d-1) \wp^{(\ell)}\cdot\wp^{(\ell)}\right),
\ee
which describes the dynamics of a countably infinite set of symmetric traceless and decoupled point-tensors. It now
remains to carry out the analysis of the variational problem for this action.

We start again from the space of asymptotic solutions. The equation of motion for the symmetric tensor of rank $\ell$ is
\be
\ddot\wp^{(\ell)}_{a_1a_2\ldots a_\ell}+\frac dr \dot\wp^{(\ell)}_{a_1a_2\ldots a_\ell}
-\frac{\ell(\ell+d-1)}{r^2}\wp^{(\ell)}_{a_1a_2\ldots a_\ell}=0,
\ee
whose general solution is
\be
\wp^{(\ell)}_{a_1a_2\ldots a_\ell}(r)=r^\ell \wp^{(\ell)}_{+a_1a_2\ldots a_\ell}
+r^{-(\ell+d-1)}\wp^{(\ell)}_{-a_1a_2\ldots a_\ell},
\ee
where $\wp^{(\ell)}_{\pm a_1a_2\ldots a_\ell}$ are arbitrary constant totally symmetric and traceless tensors.
The canonical momentum conjugate to $\wp^{(\ell)}_{a_1a_2\ldots a_\ell}$ is
\be
\pi^{(\ell)a_1a_2\ldots a_\ell}=\frac{2\p^{(d+1)/2}}{2^\ell\G\left(\frac{d+1}{2}+\ell\right)}e^{dA(r)}
\frac{1}{\ell !}\dot\wp^{(\ell)a_1a_2\ldots a_\ell},
\ee
and the symplectic form on phase space takes the form
\be
\Om_\cp= \sum_{\ell=0}^\infty\d\pi^{(\ell)a_1a_2\ldots a_\ell}\wedge \d\wp^{(\ell)}_{a_1a_2\ldots a_\ell}.
\ee
Inserting the above general solution in this symplectic form we obtain the pullback of the symplectic form
on the space of asymptotic (in this case exact) solutions, namely
\be
\Om_\cc=\vf_r^*\Om_\cp=\sum_{\ell=0}^\infty \frac{2\p^{(d+1)/2}}{2^\ell\G\left(\frac{d+1}{2}+\ell\right)}
 (2\ell+d-1)\frac{1}{\ell !}\d\wp_+^{(\ell)a_1a_2\ldots a_\ell}\wedge \d\wp^{(\ell)}_{-a_1a_2\ldots a_\ell}.
\ee
We have therefore succeeded in constructing a sensible space of asymptotic solutions that inherits a non-degenerate
symplectic form from that on phase space. The next step is to construct the boundary term that induces the
canonical transformation which makes the variational problem well defined.

The Hamilton-Jacobi equation for the tensor of rank $\ell$ takes the form
\be
\frac12 e^{dA(r)} \left(\frac{2\p^{(d+1)/2}}{2^\ell\G\left(\frac{d+1}{2}+\ell\right)}\frac{1}{\ell !}e^{dA(r)}\right)^{-2}\p^{(\ell)}\cdot\p^{(\ell)}
-\frac12 e^{(d-2)A(r)}\ell(\ell+d-1)\wp^{(\ell)}\cdot\wp^{(\ell)}+\frac{\pa\cs^{(\ell)}}{\pa r}=0,
\ee
where
\be
\pi^{(\ell)a_1a_2\ldots a_\ell}=\frac{\d\cs^{(\ell)}}{\d \wp^{(\ell)}_{a_1a_2\ldots a_\ell}}.
\ee
Inserting an ansatz of the form
\be
\cs^{(\ell)}=\frac12 e^{(d-1)A(r)} \l(r)\wp^{(\ell)}\cdot\wp^{(\ell)},
\ee
in the Hamilton-Jacobi equation, we see that $\l$ must satisfy
\be
\l^2+(d-1)\l+r\dot\l-\ell(\ell+d-1)=0.
\ee
The general solution is
\be
\l=\frac{\ell r^{d-1+2\ell}-c(d-1+\ell)}{r^{d-1+2\ell}+c},
\ee
where $c$ is an arbitrary constant. As we have seen a number of times by now, the constant $c$
interpolates between the two modes. Namely for $c=0$ we have $\l=\ell$, while for $|c|=\infty$ we get instead
$\l=-(d-1+\ell)$. Again, any $|c|<\infty$ is equally good in making the variational problem well defined, but as we will
see the only value that diagonalizes the symplectic map $\vf^*_r$ is $c=0$. We therefore take
\be\label{massless-scalar-bterm}
S_b=-\frac12 e^{(d-1)A(r)}\sum_{\ell=1}^\infty\ell\wp^{(\ell)}\cdot\wp^{(\ell)}.
\ee
It is interesting to note that this boundary term is exactly what one obtains by taking the
limit $m\to0$ in the boundary term (\ref{flat-boundary-term}) and subsequently expressing the
scalar field in terms of the traceless symmetric tensors. However, it is now clear that the transformation
induced by this boundary term preserves the symplectic form and hence it is a canonical transformation.

Finally, evaluating the map $\vf_r^*:\ct^*\cp\to \ct^*\cc$ after the canonical transformation induced by
this boundary term, we find
\be
\vf^*_r\left(\begin{matrix}
              \d \P^{(\ell)a_1a_2\ldots a_\ell}\\
              \d\wp^{(\ell)}_{a_1a_2\ldots a_\ell}
             \end{matrix}
\right)=
\left(\begin{matrix}
       -\frac{2\p^{(d+1)/2}}{2^\ell\G\left(\frac{d+1}{2}+\ell\right)}
 (2\ell+d-1)\frac{1}{\ell !}r^{-\ell}\d\wp_-^{(\ell)a_1a_2\ldots a_\ell} \\
       r^\ell \d\wp^{(\ell)}_{+a_1a_2\ldots a_\ell}+\cdots
      \end{matrix}
\right).
\ee
So, again, we see that the above boundary term diagonalizes the symplectic map $\vf_r^*:\ct^*\cp\to \ct^*\cc$.

\subsection{Einstein-Hilbert gravity and constrained phase space}
\setcounter{equation}{0}

So far we have carried out the analysis of the variational problem for a number of examples, all of which had some
distinct interesting aspects. Gravity presents us with yet another feature which we have not seen so far in the examples
we considered. Namely, the phase space description of gravity in an ADM like formalism is gauge redundant and
so the coordinates on phase space are subject to constraints. The physical phase space is the quotient,
$\cp/C$, of the unconstrained phase space space, $\cp$, by the constraints $C$. Similarly, the physical space of asymptotic
solutions is the quotient $\cc/C$ of the space of asymptotic solutions, $\cc$, by the constraints.
This is the same as the space of asymptotic solutions modulo bulk diffeomorphisms that preserve the form of the asymptotic solutions. In this subsection we will discuss how the picture of canonical transformations applies to the constrained phase
space of asymptotically (Euclidean) AdS gravity.

\subsubsection{Phase space, diffeomorphisms, and the algebra of constraints}

In order to formulate Einstein gravity in Hamiltonian language we use an ADM like formalism \cite{Arnowitt:1960es}, but with Hamiltonian time being the coordinate $r$, emanating from the boundary at $r=\infty$. This formalism applies irrespectively
of the asymptotics of spacetime. The problem we are interested in, however, requires that the hypersurfaces $\S_r$ admit a
sensible limit $r\to\infty$, which we will call the boundary, $\pa\cm$. The boundary is therefore always a codimension one
manifold. Note that this does not necessarily mean that the induced metric on $\S_r$ does not become degenerate as $r\to\infty$.
As we have explicitly seen in the previous section, the type of degrees of freedom that parameterize the space of asymptotic
solutions is not always the same as that of the bulk degrees of freedom. Hence, the degrees of freedom that parameterize
the space of asymptotic solutions of Einstein-Hilbert gravity need not be a boundary metric and its conjugate stress tensor.
Indeed, this seems to be the case for gravity in anything but asymptotically AdS spaces. Moreover,
in the case of Lorentzian bulk gravity, the boundary can be timelike, as is the case in anti de Sitter space, spacelike,
as is the case for de Sitter space, or it can even have different signature in different regions, as is the case in
Minkowski spacetime \cite{Papadimitriou-AF}.

We start with the Einstein-Hilbert action
\be\label{EH-action}
S=-\frac{1}{2\kappa^2}\int_\cm d^{d+1}x\sqrt{g}(R-2\L)-\frac{1}{2\k^2}\int_{\pa\cm}d^dx\sqrt{\g}2K,
\ee
where $\kappa^2=8\pi G_{d+1}$ is the gravitational constant, and we have included the standard
Gibbons-Hawking term \cite{Gibbons:1976ue}. This term makes the variational problem well defined in a space $\cm$ of
{\em finite} volume and it is necessary in order to formulate the dynamics in Hamiltonian language.
In the Hamiltonian formulation the  metric $g_{\m\n}$ is replaced by a set of fields $\{N,N_i,\g_{ij}\}$ on $\S_r$,
by writing
\be\label{ADM-metric}
ds^2=(N^2+N_iN^i)dr^2+2N_idrdx^i+\g_{ij}dx^idx^j,
\ee
where $N$ and  $N_i$ are respectively the lapse and shift functions,
and $\g_{ij}$ is the induced metric on the hypersurfaces $\S_r$
of constant radial coordinate $r$. In terms of these variables the Ricci scalar takes the form
\be
R[g]=R[\g]+K^2-K_{ij}K^{ij}+\nabla_\m (-2Kn^\m+n^\r\nabla_\r n^\m),
\ee
where $R[\g]$ is the Ricci scalar of the induced metric $\g_{ij}$, the extrinsic curvature, $K_{ij}$, of the hypersurface
$\S_r$ is given by
\be\label{extr-curv}
K_{ij}=\frac{1}{2N}\left(\dot{\g}_{ij}-D_iN_j-D_jN_i\right),
\ee
and $D_i$ is the covariant derivative w.r.t. the induced metric $\g_{ij}$. Moreover, $K=\g^{ij} K_{ij}$ and
$n^\m=\left(1/N,-N^i/N\right)$, is the unit normal vector to $\S_r$. The total derivative term
in this decomposition of the bulk Ricci scalar is an indication of the need for the Gibbons-Hawking term. Evaluating
this term on $\S_r$ we see that it gives a contribution which is precisely canceled by the Gibbons-Hawking term. We
therefore arrive at a Lagrangian description of the dynamics of the induced fields $\{N,N_i,\g_{ij}\}$ on $\S_r$, namely
\be\label{lagrangian-gf}
L=-\frac{1}{2\k^2}\int_{\S_r}d^dx\sqrt{\g}N\left(R[\g]-2\L+K^2-K^i_jK^j_i\right).
\ee
This Lagrangian involves no kinetic terms for the fields $N$ and $N_i$, which are therefore Lagrange multipliers, leading
to constraints.

We then proceed in the standard way by introducing the canonical momentum conjugate to $\g_{ij}$\footnote{It may be worth
emphasizing at this point that we are working within the second order formalism for Einstein gravity, where
the metric is the only independent variable. This is the appropriate formulation of gravity that arises in the
AdS/CFT correspondence. AdS gravity in the first order Palatini formalism has been discussed e.g. in
\cite{Olea:2006vd,Mansi:2008br} and boundary counterterms, so called `Kounterterms', have been derived. However, the
variational problems of first order and second order gravity are in general {\em not} the same, and correspondingly the
boundary terms required to make them well defined are generically different. In particular, since in the Palatini formalism
both the metric and the connection are independent fields, the induced metric and the extrinsic curvature on a radial slice are
independent coordinates in phase space and {\em not} canonically conjugate variables as is the case in second order gravity.
However, since the variational problem of first order gravity is more general than the second order one, the Kounterterms
\cite{Olea:2006vd}, which depend explicitly both on the extrinsic curvature and the induced metric, should reduce
to the usual counterterms of second order gravity once the relation $K_{ij}[\g]$, determined by solving the Hamilton-Jacobi
equation of second order gravity, is substituted in.}

\be\label{momentum}
\pi^{ij} =\frac{\d L}{\d\dot{\g}_{ij}}=  -\frac{1}{2 \k^2}\sqrt{\g} (K\g^{ij}-K^{ij}),
\ee
and the Hamiltonian
\be\label{hamiltonian}
H=\int_{\S_r} d^dx \p^{ij}\dot{\g}_{ij}-L=\int_{\S_r} d^dx\left(N\ch+N_i\ch^i\right),
\ee
where
\be
\ch= 2\k^2\g^{-\frac12}\left(\p^i_j\p^j_i-\frac{1}{d-1}\p^2\right)+\frac{1}{2\k^2}\sqrt{\g}\left(R[\g]-2\L\right), \quad
\ch^i= -2D_j\p^{ij}.
\ee
Hamilton's equations for the auxiliary fields $N$ and $N_i$ lead respectively to the Hamiltonian and momentum constraints
\be\label{constraints}
\ch=0,\quad \ch^i=0.
\ee

The physical phase space, therefore, is parameterized by the canonical variables $(\p^{ij},\g_{ij})$ subject to
the constraints (\ref{constraints}). In order to exhibit the significance of these constraints, we note that the symplectic form
on phase space
\be\label{symp-form-1}
\Om=\int_{\S_r}d^dx\d\p^{ij}\wedge\d\g_{ij},
\ee
leads to the Poisson bracket
\be\label{Poisson-algebra}
\left\{\g_{ij}(r,x),\p^{kl}(r,x')\right\}=\d^{(k}_i\d^{l)}_j\d^{(d)}(x-x'),
\ee
which can be realized as
\be\label{PB}
\left\{A[\g,\p],B[\g,\p]\right\}\equiv\int_{\S_r}d^dx \left(\frac{\d A}{\d\g_{ij}}\frac{\d B}{\d\p^{ij}}
-\frac{\d B}{\d\g_{ij}}\frac{\d A}{\d\p^{ij}}\right).
\ee
It follows that for any phase space function $F[\g,\p]$
\be
\left\{F[\g,\p],\g_{ij}(r,x')\right\}=-\frac{\d F[\g,\p]}{\d \p^{ij}(r,x')},\quad
\left\{F[\g,\p],\p^{ij}(r,x')\right\}=\frac{\d F[\g,\p]}{\d \g_{ij}(r,x')}.
\ee
In particular, Hamilton's equations read
\be\label{ham-eqs}
\dot{\g}_{ij}=\frac{\d H}{\d\p^{ij}}=-\left\{H,\g_{ij}\right\},\quad \dot{\p}^{ij}=-\frac{\d H}{\d\g_{ij}}=-\left\{H,\p^{ij}\right\}.
\ee

Let us now define the phase space function
\be
C[\x]=\int_{\S_r}d^dx\left(\x\ch+\x^i\ch_i\right),
\ee
where $\x(r,x)$ is an arbitrary scalar function and $\x^i(r,x)$ is an arbitrary transverse vector. Computing the
Poisson bracket of $C[\x]$ with the induced metric and its conjugate momentum we obtain
\be\label{PB-diffeo}
\left\{C[\x],\g_{ij}\right\}=\d_{\tilde\x}\g_{ij},\quad
\left\{C[\x],\p^{ij}\right\}=\d_{\tilde\x}\p^{ij},
\ee
where
\be
\tilde{\x}^\m=\left(\x/N,\x^i-\x N^i/N\right),
\ee
and the expressions on the RHS of (\ref{PB-diffeo}) stand respectively for the transformation of the induced
metric and of the momentum under the bulk diffeomorphism $x^\m\to x^\m+\tilde{\x}^\m$. This transformation
can be computed independently, i.e. without the use of the Poisson bracket, from the transformation of
the bulk metric $g_{\m\n}$ and the Christoffel symbol $\G^\m_{\r\s}$ under such a diffeomorphism. Setting the
auxiliary fields to $N=1$ and $N^i=0$, the constraint function $C[\x]$ is the generating function of a bulk diffeomorphism with parameter $\x^\m$.

Moreover, computing the Poisson bracket of the constraint function $C[\x]$ with itself, we obtain
\be\label{constraint-algebra}
\left\{C[\x],C[\x']\right\}=C[\x''],
\ee
where
\be
\x''^\m=\left(\x^i\pa_i\x'-\x'^i\pa_i\x,\quad \x^i\pa_i\x'^j-\x'^i\pa_i\x^j-(\x D^j\x'-\x'D^j\x)\right).
\ee
Hence, the constraints close onto themselves under the Poisson bracket and are therefore first class constraints.
The corresponding algebra is the algebra of bulk diffeomorphisms. A well known difficulty with this algebra is that
the structure constants are field dependent, which can be seen from the fact that the gauge parameter
of the RHS of (\ref{constraint-algebra}) is field-dependent unless $\x$ or $\x'$ vanish. We conclude that the physical phase
space, obtained from the unconstrained phase space $\cp$ by imposing the constraints (\ref{constraints}), corresponds to the
quotient $\cp/{\rm Diff}(\cm)$.

\subsubsection{Space of asymptotic solutions, asymptotic diffeos, and the algebra of constraints}

The Hamiltonian analysis of the previous subsection applies to any Einstein-Hilbert gravity. We now want to
construct the space of asymptotic solutions of Einstein's equations and see how the symplectic form
of the constrained phase space descends to a non degenerate symplectic form on this space. We will
consider only asymptotically locally AdS asymptotics here, where all the technical results are already
known in the literature. The analogous construction for asymptotically dS solutions is essentially identical
after a Wick rotation. The analysis of the asymptotically flat solutions will appear elsewhere \cite{Papadimitriou-AF}.

In the asymptotically AdS case, the space of asymptotic solutions has been known for a long time and is provided by the
Fefferman-Graham expansion \cite{FG}. Setting the lapse and shift functions to $N=1$ and $N^i=0$ as above,
the Fefferman-Graham expansion is an expansion for the induced metric $\g_{ij}$ and takes the form (setting the
AdS radius to one)
\be\label{FG}
\g_{ij}(r,x)=e^{2r}\left(g\sub{0}_{ij}(x)+e^{-2r}g\sub{2}_{ij}(x)+\cdots
+ e^{-dr}\left(-2r h\sub{d}_{ij}(x)+g\sub{d}_{ij}(x)\right)+\cdots\right),
\ee
where the term $h\sub{d}_{ij}(x)$ is non vanishing only for even boundary dimension $d$. Inserting this
asymptotic expansion in Einstein's equation with a negative cosmological constant one finds that the boundary
metric $g\sub{0}_{ij}(x)$ is left totally unconstrained while only the trace and covariant divergence of
$g\sub{d}_{ij}(x)$ are determined locally in terms of $g\sub{0}_{ij}(x)$ and its derivatives. Moreover, the
terms $g\sub{n}_{ij}(x)$, $0<n<d$, as well as $h\sub{d}_{ij}(x)$, are all locally determined in terms of
$g\sub{0}_{ij}(x)$ and its derivatives, while the terms $g\sub{n}_{ij}(x)$, $n>d$ are uniquely determined
in terms of both $g\sub{0}_{ij}(x)$ and $g\sub{d}_{ij}(x)$. Explicit expressions for the terms
$g\sub{n}_{ij}(x)$, $0<n<d$, and $h\sub{d}_{ij}(x)$ for various dimensions can be found in \cite{de_Haro:2000xn}.
Moreover, the constraints that $g\sub{d}_{ij}(x)$ satisfies take the form
\be\label{constraints-on-solutions}
D\sub{0}^i\ct_{ij}(x)=0, \quad \ct^i_i(x)=\ca(x),
\ee
where the symmetric tensor $\ct_{ij}$ is given by\footnote{This tensor is related to the tensor $t_{ij}$ in
 \cite{de_Haro:2000xn} and the renormalized canonical momentum, $\p\sub{d}^{ij}$, introduced in \cite{Papadimitriou:2005ii}.
 The precise relations are $\ct_{ij}=\frac{d}{2\k^2}t_{ij}=-2\p\sub{d}_{ij}$.}
\be
\ct_{ij}=\frac{d}{2\k^2}\left(g\sub{d}_{ij}-g\sub{0}^{kl}g\sub{d}_{kl}g\sub{0}_{ij}\right)+X_{ij}[g\sub{0}],
\ee
with $X_{ij}[g\sub{0}]$ a local function of the boundary metric and its derivatives which depends on the
dimension $d$. In particular, $X_{ij}[g\sub{0}]$ vanishes identically for odd $d$. For even $d$ it is
uniquely determined up to a multiple of $h\sub{d}_{ij}$, which corresponds to scheme dependence \cite{de_Haro:2000xn}.
For $d=2$ the scheme independent part of $X_{ij}[g\sub{0}]$ vanishes, while the expressions for $d=4$ and $d=6$ can
be found in \cite{de_Haro:2000xn}. In the AdS/CFT dictionary the tensor $\ct_{ij}$ is identified as the stress
tensor of the dual CFT. Moreover, $D\sub{0}_i$ in (\ref{constraints-on-solutions}) denotes the covariant
derivative w.r.t. the boundary metric $g\sub{0}_{ij}$, and $\ca(x)$ is the conformal anomaly \cite{Henningson:1998gx},
which again vanishes for odd $d$ and it is a local function of the boundary metric and its derivatives for
even $d$. Explicit expressions can be found in \cite{Henningson:1998gx, de_Haro:2000xn}.

Having constructed the space of asymptotic solutions, $\cc$, we are now in a position to evaluate the symplectic form
on this space. The result is
\be
\Om_\cc=\vf_r^*\Om_\cp=\int d^dx \d\p\sub{d}^{ij}\wedge \d g\sub{0}_{ij},
\ee
where\footnote{This differs from $\p\sub{d}_{ij}$ in \cite{Papadimitriou:2005ii} only in that here $\p\sub{d}_{ij}$ is defined
as a tensor density by including $\sqrt{g\sub{0}}$ in its definition.} $\p\sub{d}^{ij}\equiv -\frac12\sqrt{g\sub{0}}\ct^{ij}$.
The space of asymptotic solutions, $\cc$, therefore, is parametrized in terms of the canonical variables $\p\sub{d}^{ij}$
and $g\sub{0}_{ij}$ and it inherits a well defined symplectic form from the one on phase space. However, the canonical
variables parameterizing the space of asymptotic solutions are still constrained as they must satisfy the conditions
(\ref{constraints-on-solutions}). As we have seen this indicates that there must be some gauge redundancy in the
description of the space of asymptotic solutions. In order to exhibit and understand this gauge redundancy we proceed
as above and we introduce the Poisson bracket
\be\label{reduced-PB}
\left\{g\sub{0}_{ij}(x),\p\sub{d}^{kl}(x')\right\}=\d^{(k}_i\d^{l)}_j\d^{(d)}(x-x'),
\ee
which can be realized as
\be
\left\{A[g\sub{0},\p\sub{d}],B[g\sub{0},\p\sub{d}]\right\}=
\int d^dx\left(\frac{\d A}{\d g\sub{0}_{ij}}\frac{\d B}{\d\p\sub{d}^{ij}}-
\frac{\d B}{\d g\sub{0}_{ij}}\frac{\d A}{\d\p\sub{d}^{ij}}\right).
\ee
In order to understand the meaning of the constraints (\ref{constraints-on-solutions}) we now simply have to
evaluate the Poisson bracket of the constraint function
\be\label{reduced-constraint-function}
C[\x_o,\s]=\int d^dx \sqrt{g\sub{0}}\left(\x_o^i(x)D\sub{0}^j\ct_{ij}+\s(x)\left(\ct^i_i-\ca\right)\right),
\ee
where $\x_o^i(x)$ and $\s(x)$ are respectively an arbitrary vector field and an arbitrary scalar function on $\pa\cm$,
with the canonical fields $g\sub{0}_{ij}$ and $\p\sub{d}^{ij}$ parameterizing the space of asymptotic solutions.

Before we write down the answer, however, let us digress for a moment and try to identify the gauge redundancy
in the asymptotic expansion (\ref{FG}). Under a general bulk diffeomorphism $x^\m\to x^\m+\x^\m$ the metric
transforms as
\be
\d_\x g_{\m\n}=-\cl_\x g_{\m\n}=-\nabla_\m\x_\n-\nabla_\n\x_\m.
\ee
The subset of bulk diffeomorphisms that preserve the gauge choice $N=1$, $N^i=0$, therefore, consists of
diffeomorphisms that satisfy
\bea
&&\d_\x g_{rr}=-\cl_\x g_{rr}=-2\dot{\x}^r=0,\NO\\
&&\d_\x g_{ri}=-\cl_\x g_{ri}=-\g_{ij}(\dot{\x}^j+\pa^j\x^r)=0.
\eea
Solving these conditions we obtain
\bea\label{vectors}
&&\x^r=-\s(x),\NO\\
&&\x^i=\x_o^i(x)-\pa_j\s(x)\int_r^\infty dr'\g^{ji}(r',x).
\eea
where $\s(x)$ and $\x_o^i(x)$ are arbitrary. Inserting the expansion (\ref{FG}) into the second of these expressions gives
\be
\x^i=\x_o^i(x)-\frac{1}{2}e^{-2r}\left(g\sub{0}^{ij}-\frac12e^{-2r}g\sub{2}^{ij}+\co(e^{-4r})\right)\pa_j\s(x),
\ee
where the indices are raised or lowered using $g\sub{0}_{ij}$. Bulk diffeomorphisms of this form preserve the asymptotic
form of the Fefferman-Graham expansion (\ref{FG}). The subset of these diffeomorphisms with $\x_o^i$ vanishing are known
in the literature \cite{Imbimbo:1999bj} as Penrose-Brown-Henneaux transformations
\cite{trove.nla.gov.au/work/18287469,Brown:1986nw}. The coefficients in the Fefferman-Graham expansion, however, transform
under such diffeomorphisms, with the transformations being determined from the relation
\be
\d_\x \g_{ij}=-\cl_\x g_{ij}=-\left(L_\x\g_{ij}+2K_{ij}\x^r\right)
=-\left(D_i\x_j+D_j\x_i-2K_{ij}\s\right).
\ee
In particular, the transformation of the leading term in the Fefferman-Graham expansion transforms as
\be
\d_\x g\sub{0}_{ij}=-\left(D\sub{0}_i\x_{oj}+D\sub{0}_j\x_{oi}\right)+2\s g\sub{0}_{ij}.
\ee
Hence, $\x^i_o(x)$ corresponds to the infinitesimal parameter of a boundary diffeomorphism, while $\s(x)$
is the infinitesimal parameter of a boundary Weyl transformation. Moreover, one finds that the canonically
conjugate variable $\p\sub{d}^{ij}$ transforms as
\bea\label{momentum-trans}
\d_\x\p\sub{d}^{ij}&=&-\left(D\sub{0}_k\left(\p\sub{d}^{ij}\x_o^k\right)-\p\sub{d}^{ik}D\sub{0}_k\x_o^j
-\p\sub{d}^{jk}D\sub{0}_k\x_o^i\right)\NO\\
&&-2\s(x)\p\sub{d}^{ij}-\frac{\d }{\d g\sub{0}_{ij}}\int d^dx \sqrt{g\sub{0}}\ca\s.
\eea
The part of the transformation proportional to $\x_o^i(x)$ simply means that $\p\sub{d}^{ij}$ transforms as
a covariant density under boundary diffeomorphisms, while the transformation under boundary Weyl rescalings
follows from the relation of $\p\sub{d}^{ij}$ to the renormalized action, which we will derive in the next
subsection. Although this general form of the Weyl transformation of the renormalized stress tensor in terms
of the conformal anomaly must surely be known, we are not aware of any other place in the literature where it has appeared.
Explicit expressions for the transformation of the renormalized stress tensor under Weyl rescaling of the boundary metric for
$d=2$ and $d=4$ can be found in \cite{Imbimbo:1999bj,de_Haro:2000xn}.

It should now be no surprise that evaluating the Poisson bracket of the constraint function (\ref{reduced-constraint-function})
with the canonical fields we obtain
\be\label{PB-little-diffeos}
\left\{C[\x_o,\s],g\sub{0}_{ij}(x)\right\}=\d_\x g\sub{0}_{ij}(x),\quad
\left\{C[\x_o,\s],\p\sub{d}^{ij}(x)\right\}=\d_\x \p\sub{d}^{ij}(x).
\ee
We have therefore confirmed that the constraints (\ref{constraints-on-solutions}) generate bulk diffeomorphisms
that preserve the form of the asymptotic solution (\ref{FG}). Moreover, evaluating the Poisson bracket of the
constraint function (\ref{reduced-constraint-function}) with itself we find that the constraints close on themselves, namely
\be\label{reduced-constraint-algebra}
\left\{C[\x_o,\s],C[\x_o',\s']\right\}=C[\x_o'',\s''],
\ee
where
\be
\x_o''^i=\x_o^j\pa_j\x_o'^i-\x_o'^j\pa_j\x_o^i,\quad \s''=\x_o^j\pa_j\s'-\x_o'^j\pa_j\s.
\ee
Contrary to the phase space algebra of constraints, however, here $\x_o''^i$ and $\s''$ are not field dependent and so
the algebra of constraints is now strictly a gauge algebra. So, the physical space of asymptotic solutions is the
quotient
\be
\cc/{\rm Diff}_o(\cm),
\ee
of the space of asymptotic solutions, which are parameterized by $g\sub{0}_{ij}$ and $\p\sub{d}^{ij}$,
by the group of bulk diffeomorphisms that preserve the form of the asymptotic solutions. Note that
the reason why $\p\sub{d}^{ij}$ appears when one evaluates the symplectic form on the space of asymptotic
solutions instead of $g\sub{d}_{ij}$ is precisely because the former lifts to a good coordinate on
$\cc/{\rm Diff}_o(\cm)$, whereas the latter does not.

\subsubsection{Boundary terms and canonical transformations}

The constraints (\ref{constraints}) play another important role in theories with diffeomorphism invariance. Diffeomorphism
invariance means that Hamilton's principal function does not depend explicitly on the `time', $r$, and hence, the
Hamilton-Jacobi equation takes the form
\be
H=0,
\ee
where the Hamiltonian is given by (\ref{hamiltonian}). Since this must hold for any value of the shift and lapse
functions, the Hamilton-Jacobi equation translates to the condition that the constraints vanish, namely
\bea
&&2\k^2\g^{-\frac12}\left(\g_{ik}\g_{jl}-\frac{1}{d-1}\g_{ij}\g_{kl}\right)\frac{1}{\sqrt{\g}}\frac{\d\cs}{\d\g_{ij}}
\frac{1}{\sqrt{\g}}\frac{\d\cs}{\d\g_{ij}}=-\frac{1}{2\k^2}\sqrt{\g}\left(R[\g]-2\L\right), \NO\\
&&D_j\left(\frac{1}{\sqrt{\g}}\frac{\d\cs}{\d\g_{ij}}\right)=0.
\eea
These equations can be solved systematically in a number of different ways in order to obtain the required boundary term, $S_b$, that makes the variational problem well defined. The explicit form of the appropriate term for various dimensions can be found in the literature \cite{Henningson:1998gx, Balasubramanian:1999re,
Emparan:1999pm, Kraus:1999di, de_Boer:1999xf,de_Haro:2000xn,Bianchi:2001kw,Martelli:2002sp,Papadimitriou:2004ap}, and
there is no need to reproduce it here.

The point we want to stress here is the fact that the shift
\be
\P^{ij}=\p^{ij}+\frac{\d S_b}{\d\g_{ij}},
\ee
in the canonical momentum induced by the boundary term preserves the symplectic form, since
\bea
&&\int d^dx \d\left(\frac{\d S_b}{\d\g_{ij}}\right)\wedge \d\g_{ij}=
\int d^dx \d_1\left(\frac{\d S_b}{\d\g_{ij}}\right)\d_2\g_{ij}-(1\leftrightarrow 2)=\NO\\
&&\d_1\int d^dx \frac{\d S_b}{\d\g_{ij}}\d_2\g_{ij}-(1\leftrightarrow 2)=(\d_1\d_2-1\leftrightarrow2)S_b=0,
\eea
and hence
\be
\Om_\cp=\int d^dx \d\P^{ij}\wedge \d\g_{ij}=\int d^dx \d\p^{ij}\wedge \d\g_{ij}.
\ee
So, again, the addition of the boundary term $S_b$ amounts to a canonical transformation. Moreover, as can seen
by explicit calculation using the form of $S_b$ (this is implicitly demonstrated for example in
\cite{de_Haro:2000xn} and more directly in \cite{Papadimitriou:2004ap}), this canonical
transformation diagonalizes the symplectic map $\vf^*_r:\ct^*\cp\to\ct^*\cc$
\be
\vf^*_r\left(\begin{matrix}
              \d \P^{ij}\\
              \d\g_{ij}
             \end{matrix}
\right)=
\left(\begin{matrix}
       e^{-2r}\d\p\sub{d}^{ij}+\cdots\\
       e^{2r}\d g\sub{0}_{ij}+\cdots
      \end{matrix}
\right),
\ee
which lifts to a map between the physical spaces $\vf^*_r:\ct^*(\cp/{\rm Diff}(\cm))\to\ct^*(\cc/{\rm Diff}_o(\cm))$.
This implies in particular that
\be
\p\sub{d}^{ij}=\frac{S_{ren}}{\d g\sub{0}_{ij}},
\ee
where $S_{ren}\equiv \lim_{r\to\infty} (\cs+S_b)$. This limit exists because $S_b$ has been constructed precisely such that
$\lim_{r\to\infty} \frac{d}{dr}(\cs+S_b)=0$. This relation allows us to justify the general form of the transformation
of $\p\sub{d}^{ij}$ under boundary Weyl rescalings which we gave above. Namely,
\be
\d_\s S_{ren}=\int d^dx \p\sub{d}^{ij}\d_\s g\sub{0}_{ij}=-\int d^dx \sqrt{g\sub{0}}\ca \s,
\ee
and, therefore,
\be
\d_\s\p\sub{d}^{ij}=\d_\s\left(\frac{\d S_{ren}}{\d g\sub{0}_{ij}}\right)=-2\s \p\sub{d}^{ij}
+\frac{\d}{\d g\sub{0}_{ij}}\d_\s S_{ren},
\ee
thus justifying (\ref{momentum-trans}).

\subsection{Holographic renormalization on the string worldsheet}
\setcounter{equation}{0}

It might be somewhat surprising, but the procedure of holographic renormalization described here is in fact
carried out on p. 67 of Polchinski's first volume on string theory \cite{Polchinski:1998rq}. In the context of
2-dimensional conformal field theory, holographic renormalization amounts to constructing the Schr\"odinger
representation of the identity operator via the state-operator correspondence. As is seen from the argument of
Polchinski, and as we have argued is the case in general for holographic renormalization, constructing this
state in the WKB approximation suffices.

If the connection with holographic renormalization is not obvious, let us quickly go through Polchinski's
argument in our language. According to the state-operator correspondence, the Schr\"o-dinger
representation of the state corresponding to an operator $\ca$ is obtained by evaluating the
path integral over the interior of the unit disk with the operator $\ca$ inserted at the origin and
with fixed boundary conditions on the unit circle for the fields over which the path integral is performed. Namely,
\be
\Psi_{\ca}[X_b]=\int^{X_b} \cd Xe^{-S_P[X]}\ca(0),
\ee
where $X_b(\th)$ is the value of the field $X(r,\th)$ on the unit circle, and $S_P$ is the Polyakov
action on a disc or radius $r$, which in polar coordinates takes the from
\be
S_{P}=\frac{1}{4\p\a'}\int_0^r dr' \int_0^{2\p}d\th r'\left(\dot{X}^2+r'^{-2}(\pa_\th X)^2\right).
\ee
It is worth pointing out at this point the striking similarity between the state-operator correspondence and Witten's
prescription for the AdS/CFT partition function \cite{Witten:1998qj}.

Note that we can use conformal invariance to move the boundary circle to $r=\infty$. The above action then
in fact becomes identical to that of a free massless scalar on $\mathbb{R}^2$, which we have already
analyzed in detail. To evaluate the state corresponding to the identity operator we can use the WKB
approximation of the above path integral, which, using the result (\ref{massless-scalar-bterm}) for $d=1$
takes the form
\be
\Psi_{\mathbf{1}}[X_b]\propto \exp\left(-\frac{1}{4\p\a'} \sum_{\ell=1}^\infty\ell\wp^{(\ell)}\cdot\wp^{(\ell)}\right),
\ee
where
\be
X_b(\th)=\sum_{\ell=0}^\infty \frac{1}{\ell!}\wp^{(\ell)}_{+a_1a_2\cdots a_\ell}x^{a_1}x^{a_2}\cdots x^{a_\ell},
\ee
and $x^a=(\cos\th,\sin\th)$. This is the same result as Polchinski's, except that we have evaluated
the path integral on a circle of an arbitrary radius $r$ and we have parameterized the boundary condition
$X_b(\th)$ in terms of harmonic polynomials on the circle instead of Fourier coefficients. The relation between
the two representations can be made explicit by using the properties of the Chebyshev polynomials. The renormalized
wavefunctional corresponding to the operator $\ca$ now is computed as
\be
\Psi_{\ca}^{ren}[X_b]=\frac{\Psi_{\ca}[X_b]}{\Psi_{\mathbf{1}}[X_b]}.
\ee
This quantity makes sense for any disk radius and, in particular, it admits a finite limit as $r\to\infty$. Moreover,
Since $\Psi_{\mathbf{1}}[X_b]$ is given in the WKB approximation, this prescription amounts to adding a boundary term
$S_b$ to the Polyakov action in the path integral for $\Psi_{\ca}[X_b]$.

\subsubsection{Strings in AdS}

As a final example, let us discuss closed strings in AdS propagating all the way to the boundary of AdS
at Euclidean worldsheet time $\t=\infty$. As for gravity, we decompose the worldsheet metric as
\be
ds_{\S}^2=(N^2+\g^{\s\s}N_\s^2)d\t^2+2N_\s d\t d\s+\g_{\s\s}d\s^2,
\ee
so that the Polyakov action becomes
\be
S_{P}=\frac{1}{4\p\a'}\int_{\S} d\t d\s\sqrt{\g_{\s\s}}N G_{\m\n}(X)
\left\{\frac{1}{N^2}\left(\dot{X}^\m-N^\s{X'}^\m\right)\left(\dot{X}^\n-N^\s{X'}^\n\right)+\g_{\s\s}^{-1}{X'}^\m {X'}^\n\right\},
\ee
where $\dot{}$ denotes a derivative w.r.t. $\t$ and $'$ denotes a derivative w.r.t. $\s$. The canonical momentum conjugate
to $X^\m$ then is
\be
\p_\m=\frac{1}{2\p\a'} \sqrt{\g_{\s\s}}\frac{1}{N}G_{\m\n}(X)\left(\dot{X}^\n-N^\s{X'}^\n\right),
\ee
and the Hamiltonian takes the form
\be
H=\frac12\int_{0}^{2\p}d\s\left(N\ch+N^\s\ch_\s\right),
\ee
where
\be
\ch=\frac{2\p\a'}{\sqrt{\g_{\s\s}}}G^{\m\n}(X)\p_\m\p_\n-\frac{\sqrt{\g_{\s\s}}}{2\p\a'}G_{\m\n}(X)\g^{\s\s}X'^\m X'^\n,\quad
\ch_\s=2\p_\m X'^\m.
\ee
As for gravity, Hamilton's equations for the auxiliary fields $N$ and $N^\s$ lead to the constraints
\be\label{string-constraints}
\ch=0, \quad \ch_\s=0.
\ee
These are of course the well known Virasoro constraints, which are equivalent to setting the worldsheet stress tensor to zero.
Note that although $\g_{\s\s}$ appears as yet another auxiliary field, in fact it gives rise to the same constraint as $N$. Indeed,
$\g_{\s\s}$ in the Hamiltonian can be removed by a rescaling of $N$. This is a consequence of the tracelessness of the stress
tensor, which follows from the worldsheet Weyl invariance of the Polyakov action.

So, as in Einstein-Hilbert gravity, the constraints will become the Hamilton-Jacobi equations and the symplectic
form on phase space is
\be
\Om_\cp=\int_0^{2\p}d\s \d\p_\m\wedge \d X^\m,
\ee
which translates to the Poisson bracket
\be
\left\{X^\m(\t,\s),\p_\n(\t,\s')\right\}=\d^\m_\n\d(\s-\s').
\ee
Using this Poisson bracket one can verify that the constraints (\ref{string-constraints}) generate worldsheet diffeomorphisms
and the algebra of constraints is two copies of the Virasoro
algebra with zero central charge.\footnote{In order to find the quantum Virasoro algebra with non zero central charge the phase space must be quantized. We should point out, however, that the algebra of constraints (\ref{reduced-constraint-algebra}) in the case of three dimensional gravity leads to two copies of the Virasoro algebra with {\em non} zero central charge. This is the
famous result of Brown and Henneaux \cite{Brown:1986nw}, which was probably the first hint of the AdS/CFT correspondence. In
that case, the {\em classical} phase space of three dimensional AdS gravity describes the dynamics of a {\em quantum} two dimensional CFT in the large $N$ limit.}

We now consider the string sigma model with an AdS target space. Writing the AdS metric in Poincar\'e coordinates,
\be
ds^2=G_{\m\n}(X)dX^\m dX^\n = dr^2+e^{2r}dX^idX^i,
\ee
and gauge fixing the worldsheet metric to
\be
N=e^{\om(\t,\s)},\quad N^\s=0,\quad \g_{\s\s}=e^{2\t+2\om(\t,\s)},
\ee
where $\om(\t,\s)$ is an arbitrary function, the equations of motion and constraints for the target space
fields, $r(\t,\s)$ and $X^i(\t,\s)$ become
\bea
&&e^{-\t}\pa_\t(e^{\t}\dot{r})+\pa_\s(e^{-2\t}r')
-e^{2r}\left(\dot{X}^i\dot{X}^i+e^{-2\t}{X'}^i {X'}^i\right)=0,\NO\\
&&\pa_\t\left(e^{\t}e^{2r}\dot{X}^i\right)+\pa_\s\left(e^{-\t}e^{2r}{X'}^i\right)=0,
\eea
\be
\dot{r}^2+e^{-2\t}{r'}^2+e^{2r}\left(\dot{X}^i\dot{X}^i-e^{-2\t}{X'}^i {X'}^i\right)=0,\quad
\dot{r}r'+e^{2r}\dot{X}^i{X'}^i=0.
\ee

Looking for solutions such that the string worldsheet extents to the AdS boundary at $r=\infty$ as $\t\to\infty$,
we find that the above equations admit the following general asymptotic solution
\bea\label{string-sol}
r(\t,\s)&=&\t -\frac12\log(X_o^i{}'X_o^i{}')
+\frac16 e^{-2\t}\left(\frac{3X_o^i{}''X_o^i{}''+X_o^i{}'X_o^i{}'''}{X_o^j{}'X_o^j{}'}-\left(\frac{X_o^i{}'X_o^i{}''}{X_o^j{}'X_o^j{}'}\right)^2\right)\NO\\
&&+\frac13e^{-3\t}\left(\frac{X_o^i{}''\tilde{X}^i}{X_o^j{}'X_o^j{}'}-2\frac{X_o^j{}'X_o^j{}''X_o^i{}'\tilde{X}^i}{(X_o^j{}'X_o^j{}')^2}\right)+\co(e^{-4\t}),\NO\\
X^i(\t,\s)&=&X_o^i+e^{-2\t}\left(\frac12X_o^i{}''-\frac{X_o^j{}'X_o^j{}''X_o^i{}'}{X_o^k{}'X_o^k{}'}\right)+\frac13\tilde{X}^ie^{-3\t}+\co(e^{-4\t}),
\eea
where $X_o^i(\s)$ and $\tilde{X}^i(\s)$ are only subject to the constraints
\be
\quad \tilde{X}^i(\s)X_o^i{}'(\s)=0, \quad X_o^i{}'(\s)X_o^i{}'(\s)>0.
\ee
These otherwise arbitrary functions parameterize the space of asymptotic solutions. Evaluating the pullback
of the symplectic form on this space we find
\be
\Om_\cc=\vf^*_\t\Om_\cp=-\frac{1}{2\p\a'}\int_0^{2\p}d\s \frac{\d \tilde{X}^i\wedge \d X_o^i}{X_o^j{}'(\s)X_o^j{}'(\s)},
\ee
leading to the Poisson bracket
\be
\{X_o^i(\s),\tilde{X}^j(\s')\}=2\p\a'X_o^k{}'(\s)X_o^k{}'(\s)\h^{ij}\d(\s-\s').
\ee

It is easily seen that the only worldsheet diffeomorphisms that preserve the form of the asymptotic expansion and
the gauge fixing of the worldsheet metric are translations along $\s$, i.e. $\x^a=(0,\ve)$ for some infinitesimal
constant parameter $\ve$. Moreover, using the above Poisson bracket one can verify that the constraint function
\be
C[\ve]=\int_0^{2\p} d\s \ve\frac{\tilde{X}^i(\s)X_o^i{}'(\s)}{X_o^j{}'(\s)X_o^j{}'(\s)},
\ee
indeed generates $\s$-translations. The physical space of asymptotic solutions, therefore, is the quotient
$\cc/U(1)$, of the space of asymptotic solutions by the group of the constraint algebra.

What remains now is to determine the boundary term, $S_b$, which makes the variational problem well defined. Writing
the momenta as derivatives of Hamilton's principal function, $\cs$, the constraints (\ref{string-constraints}) lead to the
Hamilton-Jacobi equations
\bea
\label{string-HJ}
&&2\p\a'\left(\p_r^2+e^{-2r}\p_i\p_i\right)+\frac{1}{2\p\a'}\left(r'^2-e^{2r}X^i{}'X^i{}'\right)=0,\NO\\
&&r'\p_r+X^i{}'\p_i=0,
\eea
where the dependence of the worldline metric $\g_{\s\s}$ cancels out. Now, from the leading asymptotic form of
our solution, we see that the first of the Hamilton-Jacobi equations to leading order becomes
\be
2\p\a' \p_r^2\approx\frac{1}{2\p\a'}e^{2r}X^i{}'X^i{}',
\ee
which can be directly integrated to give
\be
\cs=\frac{1}{2\p\a'}\int_0^{2\p}d\s e^r\sqrt{X^i{}'X^i{}'}+\tilde\cs,
\ee
where $\tilde\cs$ stands for subleading terms, and the sign has been fixed by matching to the
on-shell action. Inserting this form of Hamilton's principal function back to the full Hamilton-Jacobi
equation, one can easily show that $\tilde\cs$ admits a finite limit as $\t\to\infty$. We would
therefore like to take our boundary term to be
\be\label{string-bterm}
S_b=-\frac{1}{2\p\a'}\int_0^{2\p}d\s e^{r}\sqrt{{X'}^i {X'}^i}.
\ee
In order for this term to be admissible though, we need to show that it preserves the symplectic form. We have,
\bea
&&-2\p\a'\int_0^{2\p}d\s \left\{\d\left(\frac{\d S_b}{\d r}\right)\wedge \d r
+\d\left(\frac{\d S_b}{\d X^i}\right)\wedge \d X^i\right\}=\NO\\
&&\int_0^{2\p}d\s \left\{\frac{e^r{X'}^i }{\sqrt{{X'}^j {X'}^j}}\d{X'}^i\wedge \d r
-\d\pa_\s\left(\frac{e^r{X'}^i }{\sqrt{{X'}^j {X'}^j}}\right)\wedge \d X^i\right\}=\\
&&\int_0^{2\p}d\s \left\{\frac{e^r{X'}^i }{\sqrt{{X'}^j {X'}^j}}\d{X'}^i\wedge \d r+
\frac{e^r{X'}^i }{\sqrt{{X'}^j {X'}^j}}\d r\wedge \d {X'}^i+e^r\d\left(\frac{{X'}^i }{\sqrt{{X'}^j {X'}^j}}\right)
\wedge \d {X'}^i\right\}=0.\NO
\eea
It follows that the boundary term (\ref{string-bterm}) is the desired boundary term that makes the variational
problem well defined and preserves the symplectic form. Notice that this term is just the proper length of the
boundary loop where the string worldsheet intersects the boundary.

We have now shown that the transformation
\be
\left(\begin{matrix}
              \p_r\\
              \p_i\\
               r  \\
               X^i \\
             \end{matrix}
\right)\mapsto
\left(\begin{matrix}
              \P_r\\
              \P_i\\
               r  \\
               X^i \\
             \end{matrix}
\right):=
\left(\begin{matrix}
              \p_r+\frac{\d S_b}{\d r}\\
              \p_i+\frac{\d S_b}{\d X^i}\\
               r  \\
               X^i \\
             \end{matrix}
\right),
\ee
where $S_b$ is given by (\ref{string-bterm}), is canonical.  Using the asymptotic solution (\ref{string-sol}) we find
\be
\vf_\t^*
\left(\begin{matrix}
              \d\P_r\\
              \d\P_i\\
             \end{matrix}
\right)=
\left(\begin{matrix}
             \co(e^{-2\t}) \\
              -\frac{1}{2\p\a'}\d\left(\frac{\tilde{X}^i}{X_o^j{}'X_o^j{}'}\right)+\co(e^{-2\t})
             \end{matrix}
\right),
\ee
and so, once again, the canonical transformation that makes the variational problem well defined simultaneously
diagonalizes the symplectic map $\vf_\t^*:\ct^*(\cp/C)\to \ct^*(\cc/U(1))$. Note that the canonically transformed
momentum conjugate to $r$ vanishes as $\t\to\infty$, which reflects the fact that $r$ is not an independent dynamical
variable but is expressed in terms of $X_0^i$.

\section*{Acknowledgements}

I would like to thank Steven Gubser for a useful conversation and Thomas Konstandin for various comments on the
manuscript. I would also like to thank the organizers of the 2010 Amsterdam Workshop on String Theory, during
which most of this work was done.

\providecommand{\href}[2]{#2}\begingroup\raggedright\endgroup


\begin{thebibliography}{10}

\bibitem{Henningson:1998gx}
M.~Henningson and K.~Skenderis, {\it {The holographic Weyl anomaly}},  {\em
  JHEP} {\bf 07} (1998) 023,
  [\href{http://xxx.lanl.gov/abs/hep-th/9806087}{{\tt hep-th/9806087}}].

\bibitem{Balasubramanian:1999re}
V.~Balasubramanian and P.~Kraus, {\it {A stress tensor for anti-de Sitter
  gravity}},  {\em Commun. Math. Phys.} {\bf 208} (1999) 413--428,
  [\href{http://xxx.lanl.gov/abs/hep-th/9902121}{{\tt hep-th/9902121}}].

\bibitem{Emparan:1999pm}
R.~Emparan, C.~V. Johnson, and R.~C. Myers, {\it {Surface terms as counterterms
  in the AdS/CFT correspondence}},  {\em Phys. Rev.} {\bf D60} (1999) 104001,
  [\href{http://xxx.lanl.gov/abs/hep-th/9903238}{{\tt hep-th/9903238}}].

\bibitem{Kraus:1999di}
P.~Kraus, F.~Larsen, and R.~Siebelink, {\it {The gravitational action in
  asymptotically AdS and flat spacetimes}},  {\em Nucl. Phys.} {\bf B563}
  (1999) 259--278, [\href{http://xxx.lanl.gov/abs/hep-th/9906127}{{\tt
  hep-th/9906127}}].

\bibitem{de_Boer:1999xf}
J.~de~Boer, E.~P. Verlinde, and H.~L. Verlinde, {\it {On the holographic
  renormalization group}},  {\em JHEP} {\bf 08} (2000) 003,
  [\href{http://xxx.lanl.gov/abs/hep-th/9912012}{{\tt hep-th/9912012}}].

\bibitem{de_Haro:2000xn}
S.~de~Haro, S.~N. Solodukhin, and K.~Skenderis, {\it {Holographic
  reconstruction of spacetime and renormalization in the AdS/CFT
  correspondence}},  {\em Commun. Math. Phys.} {\bf 217} (2001) 595--622,
  [\href{http://xxx.lanl.gov/abs/hep-th/0002230}{{\tt hep-th/0002230}}].

\bibitem{Bianchi:2001kw}
M.~Bianchi, D.~Z. Freedman, and K.~Skenderis, {\it {Holographic
  Renormalization}},  {\em Nucl. Phys.} {\bf B631} (2002) 159--194,
  [\href{http://xxx.lanl.gov/abs/hep-th/0112119}{{\tt hep-th/0112119}}].

\bibitem{Martelli:2002sp}
D.~Martelli and W.~Mueck, {\it {Holographic renormalization and Ward identities
  with the Hamilton-Jacobi method}},  {\em Nucl. Phys.} {\bf B654} (2003)
  248--276, [\href{http://xxx.lanl.gov/abs/hep-th/0205061}{{\tt
  hep-th/0205061}}].

\bibitem{Papadimitriou:2004ap}
I.~Papadimitriou and K.~Skenderis, {\it {AdS / CFT correspondence and
  geometry}},  \href{http://xxx.lanl.gov/abs/hep-th/0404176}{{\tt
  hep-th/0404176}}.

\bibitem{Hartle:1983ai}
J.~Hartle and S.~Hawking, {\it {Wave Function of the Universe}},  {\em
  Phys.Rev.} {\bf D28} (1983) 2960--2975.

\bibitem{Seiberg:1990eb}
N.~Seiberg, {\it {Notes on quantum Liouville theory and quantum gravity}},
  {\em Prog.Theor.Phys.Suppl.} {\bf 102} (1990) 319--349.

\bibitem{Papadimitriou-AF}
I.~Papadimitriou, {\it {Holographic renormalization for asymptotically flat
  gravity}}. In preparation.

\bibitem{Papadimitriou-HQCD}
I.~Papadimitriou, {\it {Holographic renormalization for Improved Holographic
  QCD}}. In preparation.

\bibitem{Skenderis:2006uy}
K.~Skenderis and M.~Taylor, {\it {Kaluza-Klein holography}},  {\em JHEP} {\bf
  0605} (2006) 057, [\href{http://xxx.lanl.gov/abs/hep-th/0603016}{{\tt
  hep-th/0603016}}].

\bibitem{Reed:1975uy}
M.~Reed and B.~Simon, {\it {Methods of Modern Mathematical Physics. 2. Fourier
  Analysis, Selfadjointness}}. New York 1975, 361p.

\bibitem{Carreau:1990is}
M.~Carreau, E.~Farhi, S.~Gutmann, and P.~F. Mende, {\it {The Functional
  Integral for Quantum Systems with Hamiltonians Unbounded from Below}},  {\em
  Ann. Phys.} {\bf 204} (1990) 186--207.

\bibitem{Papadimitriou:2005ii}
I.~Papadimitriou and K.~Skenderis, {\it {Thermodynamics of asymptotically
  locally AdS spacetimes}},  {\em JHEP} {\bf 08} (2005) 004,
  [\href{http://xxx.lanl.gov/abs/hep-th/0505190}{{\tt hep-th/0505190}}].

\bibitem{Breitenlohner:1982jf}
P.~Breitenlohner and D.~Z. Freedman, {\it {Stability in Gauged Extended
  Supergravity}},  {\em Ann. Phys.} {\bf 144} (1982) 249.

\bibitem{Klebanov:1999tb}
I.~R. Klebanov and E.~Witten, {\it {AdS/CFT correspondence and symmetry
  breaking}},  {\em Nucl. Phys.} {\bf B556} (1999) 89--114,
  [\href{http://xxx.lanl.gov/abs/hep-th/9905104}{{\tt hep-th/9905104}}].

\bibitem{Witten:2001ua}
E.~Witten, {\it {Multi-trace operators, boundary conditions, and AdS/CFT
  correspondence}},  \href{http://xxx.lanl.gov/abs/hep-th/0112258}{{\tt
  hep-th/0112258}}.

\bibitem{Witten:2003ya}
E.~Witten, {\it {SL(2,Z) action on three-dimensional conformal field theories
  with Abelian symmetry}},  \href{http://xxx.lanl.gov/abs/hep-th/0307041}{{\tt
  hep-th/0307041}}.

\bibitem{Witten:1998qj}
E.~Witten, {\it {Anti-de Sitter space and holography}},  {\em Adv. Theor. Math.
  Phys.} {\bf 2} (1998) 253--291,
  [\href{http://xxx.lanl.gov/abs/hep-th/9802150}{{\tt hep-th/9802150}}].

\bibitem{Rey:1998bq}
S.-J. Rey, S.~Theisen, and J.-T. Yee, {\it {Wilson-Polyakov loop at finite
  temperature in large N gauge theory and anti-de Sitter supergravity}},  {\em
  Nucl. Phys.} {\bf B527} (1998) 171--186,
  [\href{http://xxx.lanl.gov/abs/hep-th/9803135}{{\tt hep-th/9803135}}].

\bibitem{Drukker:1999zq}
N.~Drukker, D.~J. Gross, and H.~Ooguri, {\it {Wilson loops and minimal
  surfaces}},  {\em Phys. Rev.} {\bf D60} (1999) 125006,
  [\href{http://xxx.lanl.gov/abs/hep-th/9904191}{{\tt hep-th/9904191}}].

\bibitem{Alday:2007hr}
L.~F. Alday and J.~M. Maldacena, {\it {Gluon scattering amplitudes at strong
  coupling}},  {\em JHEP} {\bf 06} (2007) 064,
  [\href{http://xxx.lanl.gov/abs/0705.0303}{{\tt 0705.0303}}].

\bibitem{Skenderis:2002wp}
K.~Skenderis, {\it {Lecture notes on holographic renormalization}},  {\em
  Class. Quant. Grav.} {\bf 19} (2002) 5849--5876,
  [\href{http://xxx.lanl.gov/abs/hep-th/0209067}{{\tt hep-th/0209067}}].

\bibitem{Lee:1990nz}
J.~Lee and R.~M. Wald, {\it {Local symmetries and constraints}},  {\em J. Math.
  Phys.} {\bf 31} (1990) 725--743.

\bibitem{Papadimitriou:2003is}
I.~Papadimitriou, {\it {Holographic renormalization made simple: An example}},
  . Prepared for International School of Subnuclear Physics: 41st Course: From
  Quarks to Black Holes: Progress in Understanding the Logic of Nature, Erice,
  Sicily, Italy, 29 Aug - 7 Sep 2003.

\bibitem{abramowitz+stegun}
M.~{Abramowitz} and I.~A. {Stegun}, {\em Handbook of Mathematical Functions
  with Formulas, Graphs, and Mathematical Tables}.
\newblock Dover, New York, ninth dover printing, tenth gpo printing~ed., 1964.

\bibitem{deBoer:2003vf}
J.~de~Boer and S.~N. Solodukhin, {\it {A Holographic reduction of Minkowski
  space-time}},  {\em Nucl.Phys.} {\bf B665} (2003) 545--593,
  [\href{http://xxx.lanl.gov/abs/hep-th/0303006}{{\tt hep-th/0303006}}].

\bibitem{Arnowitt:1960es}
R.~L. Arnowitt, S.~Deser, and C.~W. Misner, {\it {Canonical variables for
  general relativity}},  {\em Phys.Rev.} {\bf 117} (1960) 1595--1602.

\bibitem{Gibbons:1976ue}
G.~Gibbons and S.~Hawking, {\it {Action Integrals and Partition Functions in
  Quantum Gravity}},  {\em Phys.Rev.} {\bf D15} (1977) 2752--2756.

\bibitem{Olea:2006vd}
  R.~Olea,
  ``Regularization of odd-dimensional AdS gravity: Kounterterms,''
  JHEP {\bf 0704} (2007) 073
  [arXiv:hep-th/0610230].

\bibitem{Mansi:2008br}
  D.~S.~Mansi, A.~C.~Petkou and G.~Tagliabue,
  ``Gravity in the 3+1-Split Formalism I: Holography as an Initial Value Problem,''
  Class.\ Quant.\ Grav.\  {\bf 26} (2009) 045008
  [arXiv:0808.1212 [hep-th]].


\bibitem{FG}
C.~Fefferman and C.~R. Graham, {\it {Conformal Invariants}},  {\em Elie Cartan
  et les Math{\'e}matiques d'aujourd'hui, Ast{\'e}risque} (1985).

\bibitem{Imbimbo:1999bj}
C.~Imbimbo, A.~Schwimmer, S.~Theisen, and S.~Yankielowicz, {\it
  {Diffeomorphisms and holographic anomalies}},  {\em Class. Quant. Grav.} {\bf
  17} (2000) 1129--1138, [\href{http://xxx.lanl.gov/abs/hep-th/9910267}{{\tt
  hep-th/9910267}}].

\bibitem{trove.nla.gov.au/work/18287469}
R.~R. Penrose and W.~{1924- Rindler}, {\em Spinors and space-time / Roger
  Penrose, Wolfgang Rindler. - Vol.2., Spinor and twistor methods in space-time
  geometry}.
\newblock Cambridge : Cambridge University Press, 1986.

\bibitem{Brown:1986nw}
J.~Brown and M.~Henneaux, {\it {Central Charges in the Canonical Realization of
  Asymptotic Symmetries: An Example from Three-Dimensional Gravity}},  {\em
  Commun.Math.Phys.} {\bf 104} (1986) 207--226.

\bibitem{Polchinski:1998rq}
J.~Polchinski, {\em {String theory. Vol. 1: An introduction to the bosonic
  string}}.
\newblock Cambridge : Cambridge University Press, 1998.

\end{thebibliography}
\end{document}